\documentclass[journal,twoside,web]{ieeecolor}

\usepackage{fontawesome5}

\usepackage{generic}
\usepackage{cite}
\usepackage{amsmath,amssymb,amsfonts}
\usepackage{graphicx}
\usepackage{algorithm}
\usepackage{hyperref}
\hypersetup{hidelinks=true}
\usepackage{textcomp}

\usepackage{optidef}

\usepackage{tikzpeople}
\usepackage{color}
\usepackage{xspace,stackrel}
\usepackage{wrapfig}
\usepackage{mathtools}
\usepackage{tikz,pgfplots}
\usepackage{etoolbox}
\usetikzlibrary{external} 
\usepackage{autonum}
\usepackage{dsfont}

\usepackage{multirow}
\usepackage{algpseudocode}
\usetikzlibrary{arrows.meta}

\DeclareMathOperator*{\argmax}{arg\,max}

\usepgfplotslibrary{fillbetween}
\usepgfplotslibrary{groupplots}
\usetikzlibrary{decorations.pathreplacing}

\usepackage{lipsum,adjustbox}

\DeclareMathAlphabet{\mathcal}{OMS}{cmsy}{m}{n}

\def\beq{\begin{equation}}
\def\eeq{\end{equation}}

\newcommand{\mc}{\mathcal}

\newcommand{\Z}{\mathbb{Z}}

\newcommand{\R}{\mathds{R}}
\newcommand{\N}{\mathbb{N}}

\newcommand{\defineas}{\coloneqq}

\newcommand{\norm}[1]{\left\lVert#1\right\rVert}

\newtheorem{assumption}{Assumption}
\newtheorem{standing assumption}{Standing Assumption}

\newtheorem{theorem}{Theorem}
\newtheorem{proposition}{Proposition}
\newtheorem{corollary}{Corollary}
\newtheorem{definition}{Definition}
\newtheorem{lemma}{Lemma}
\newtheorem{remark}{Remark}

\newtoggle{full_version}
\toggletrue{full_version}

\definecolor{mycolor1}{RGB}{230,97,1}
\definecolor{mycolor2}{RGB}{178,171,210}
\definecolor{mycolor3}{RGB}{253,184,99}
\definecolor{mycolor4}{RGB}{94,60,153}
\definecolor{mycolor5}{rgb}{0,0,0}

\tikzset{
  pics/car/.style args={#1}{
     code={
     \begin{scope}[scale=0.15]
      \shade[top color=#1, bottom color=white, shading angle={135}]
        [draw=black,fill=red!20,rounded corners=0.2ex] (1.5,.5) -- ++(0,1) -- ++(1,0.3) --  ++(3,0) -- ++(1,0) -- ++(0,-1.3) -- (1.5,.5) -- cycle;
    \draw[ rounded corners=0.5ex,fill=black!20!blue!20!white]  (2.5,1.8) -- ++(1,0.7) -- ++(1.6,0) -- ++(0.6,-0.7) -- (2.5,1.8);
    \draw[thick]  (4.2,1.8) -- (4.2,2.5);
    \draw[draw=black,fill=gray!50,thick] (2.75,.5) circle (.5);
    \draw[draw=black,fill=gray!50,thick] (5.5,.5) circle (.5);
    \end{scope}
     }
  }
}

\def\BibTeX{{\rm B\kern-.05em{\sc i\kern-.025em b}\kern-.08em
    T\kern-.1667em\lower.7ex\hbox{E}\kern-.125emX}}
\begin{document}

\title{Resource-Splitting Games with Tullock-Based Lossy Contests}

\author{Marko Maljkovic, Gustav Nilsson, and Nikolas Geroliminis
\thanks{M.~Maljkovic and N.~Geroliminis are with the School of Architecture, Civil and Environmental Engineering, École Polytechnique Fédérale de Lausanne (EPFL), 1015 Lausanne, Switzerland. {\tt\small \{marko.maljkovicnikolas.geroliminis\}@epfl.ch}.}
\thanks{G.~Nilsson is with the Department of Industrial Engineering, University of Trento, Italy {\tt\small gustav.nilsson@unitn.it.}}
\thanks{Part of
the results of this article appeared in a preliminary form in~\cite{maljkovic2024blotto} and~\cite{maljkovic2024cdc}.}
\thanks{This work was supported by the Swiss National Science Foundation under NCCR Automation, grant agreement 51NF40\_180545.}
}

\maketitle

\begin{abstract}
This paper introduces a novel class of multi-stage resource allocation games that model real-world scenarios in which profitability depends on the balance between supply and demand, and where higher resource investment leads to greater returns. Our proposed framework, which incorporates the notion of profit loss due to insufficient player participation, gives rise to a Tullock-like functional form of the stage payoff structure when weighted fair proportional resource allocation is applied. We explore both centralized and Nash equilibrium strategies, establish sufficient conditions for their existence and uniqueness, and provide an iterative, semi-decentralized method to compute the Nash equilibrium in games with arbitrarily many players. Additionally, we demonstrate that the framework generalizes instances of several existing models, including Receding Horizon and Blotto games, and present a semi-analytical method for computing the unique Nash equilibrium within the Blotto setup. Our findings are validated through a numerical case study in smart mobility, highlighting the practical relevance and applicability of the proposed model.
\end{abstract}


\section{Introduction}
Resource-splitting games have emerged as one of the fundamental areas in game theory, focused on understanding the intricate dynamics that emerge when multiple agents compete or collaborate to allocate limited resources between them. At their core, these games revolve around the idea that each agent's strategic choices are governed by a personal objective they seek to maximize, all within the constraints of resource scarcity and established norms of behavior. Typically, players interact over multiple stages within a broader market, with each stage offering the potential for a certain profit. Naturally, the share of the market in every stage that players secure depends on the resources they invest and directly contributes to their overall profit. In this sense, the term ``stage'' is polysemic and can be understood from both spatial and temporal perspectives. Spatially distributed stages, often referred to as battlefields, relate to simultaneous strategic contests happening on multiple fronts or in multiple regions, while temporally distributed stages, known as ``time steps'', typically involve the evolution of players' internal dynamics over a sequence of time periods. Regardless of the interpretation, their flexible general structure has made these games applicable to many real-world problems in the fields of smart mobility~\cite{maljkovic2024blotto, maljkovic2024cdc}, telecommunications~\cite{Kelly,4346554,8368291}, energy~\cite{9992497,6994293, 6987341}, and operational management~\cite{NICOSIA2017933,6747393,KIM20181}. 
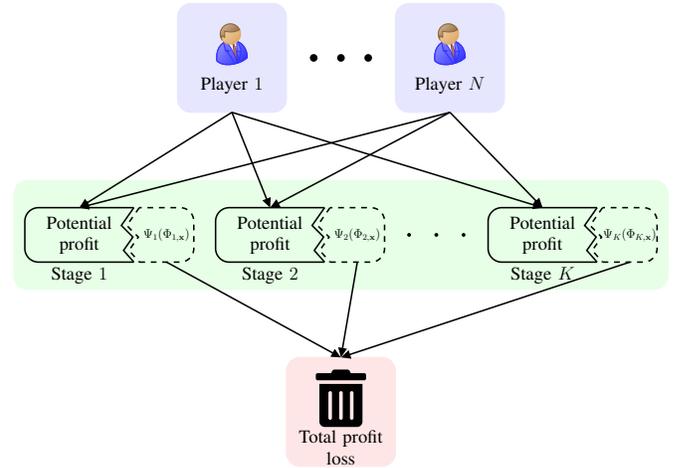
\begin {figure}
\centering
\begin{adjustbox}{max height=0.65\textwidth, max width=0.48\textwidth}
\begin{tikzpicture}[scale=1.0, font=\large]

    \coordinate (A1) at (0, 0);
    \coordinate (B1) at ($(A1)+(180:2)$);
    \coordinate (C1) at ($(B1)+(90:1)$);
    \coordinate (D1) at ($(C1)+(0:2)$);
    
    \coordinate (E1) at ($(D1)+(235:0.4)$);
    \coordinate (F1) at ($(E1)+(315:0.3)$);
    \coordinate (G1) at ($(F1)+(235:0.3)$);

    \coordinate (D11) at ($(D1)+(0:0.1)$);
    \coordinate (E11) at ($(D11)+(235:0.4)$);
    \coordinate (F11) at ($(E11)+(315:0.3)$);
    \coordinate (G11) at ($(F11)+(235:0.3)$);
    \coordinate (A11) at ($(A1)+(0:0.1)$);

    \coordinate (B11) at ($(A11)+(0:1)$);
    \coordinate (C11) at ($(B11)+(90:1)$);


    \coordinate (A2) at (3.5, 0);
    \coordinate (B2) at ($(A2)+(180:2)$);
    \coordinate (C2) at ($(B2)+(90:1)$);
    \coordinate (D2) at ($(C2)+(0:2)$);
    
    \coordinate (E2) at ($(D2)+(235:0.4)$);
    \coordinate (F2) at ($(E2)+(315:0.3)$);
    \coordinate (G2) at ($(F2)+(235:0.3)$);

    \coordinate (D21) at ($(D2)+(0:0.1)$);
    \coordinate (E21) at ($(D21)+(235:0.4)$);
    \coordinate (F21) at ($(E21)+(315:0.3)$);
    \coordinate (G21) at ($(F21)+(235:0.3)$);
    \coordinate (A21) at ($(A2)+(0:0.1)$);

    \coordinate (B21) at ($(A21)+(0:1)$);
    \coordinate (C21) at ($(B21)+(90:1)$);
    

    \coordinate (A3) at (8.5, 0);
    \coordinate (B3) at ($(A3)+(180:2)$);
    \coordinate (C3) at ($(B3)+(90:1)$);
    \coordinate (D3) at ($(C3)+(0:2)$);
    
    \coordinate (E3) at ($(D3)+(235:0.4)$);
    \coordinate (F3) at ($(E3)+(315:0.3)$);
    \coordinate (G3) at ($(F3)+(235:0.3)$);

    \coordinate (D31) at ($(D3)+(0:0.1)$);
    \coordinate (E31) at ($(D31)+(235:0.4)$);
    \coordinate (F31) at ($(E31)+(315:0.3)$);
    \coordinate (G31) at ($(F31)+(235:0.3)$);
    \coordinate (A31) at ($(A3)+(0:0.1)$);

    \coordinate (B31) at ($(A31)+(0:1)$);
    \coordinate (C31) at ($(B31)+(90:1)$);

    \coordinate (M) at ($(B1)!0.5!(B31)+(90:0.5)$);

    \node[draw=green!10, rectangle, rounded corners=2.5mm, minimum width=12cm, minimum height=2cm, anchor=center, fill=green!10] at (M) {};

        
    \draw[rounded corners=2.5mm, line width=0.8pt] (A1) -- (B1) -- (C1) -- (D1);
    \draw[line width=0.8pt] (D1) -- (E1) -- (F1) -- (G1) -- (A1);   
    \draw[dashed, line width=0.8pt] (D11) -- (E11) -- (F11) -- (G11) -- (A11);
    \draw[dashed, rounded corners=2.5mm, line width=0.8pt] (A11) -- (B11) -- (C11) -- (D11);

    \draw[rounded corners=2.5mm, line width=0.8pt] (A2) -- (B2) -- (C2) -- (D2);
    \draw[line width=0.8pt] (D2) -- (E2) -- (F2) -- (G2) -- (A2);
    \draw[dashed, line width=0.8pt] (D21) -- (E21) -- (F21) -- (G21) -- (A21);
    \draw[dashed, rounded corners=2.5mm, line width=0.8pt] (A21) -- (B21) -- (C21) -- (D21);
    
    \draw[rounded corners=2.5mm, line width=0.8pt] (A3) -- (B3) -- (C3) -- (D3);
    \draw[line width=0.8pt] (D3) -- (E3) -- (F3) -- (G3) -- (A3);
    \draw[dashed, line width=0.8pt] (D31) -- (E31) -- (F31) -- (G31) -- (A31);
    \draw[dashed, rounded corners=2.5mm, line width=0.8pt] (A31) -- (B31) -- (C31) -- (D31);


    \coordinate (ADV) at ($(B1)!0.5!(B31)+(270:2.75)$);
    \coordinate (ADV2) at ($(ADV) + (270:0.65)$);
    \coordinate (ADV3) at ($(ADV) + (90:0.25)$);
    
    \node[draw=red!10, rectangle, rounded corners=2.5mm, minimum width=2cm, minimum height=2cm, anchor=center, fill=red!10](b2) at (ADV){};
    \node[scale=0.8, align=center](b3) at (ADV2){Total profit\\loss};
    \node[ scale=2.5](b1) at (ADV3){\faTrash*};

    \coordinate (2ADV) at ($(C1)!0.5!(C31)+(90:2.75)$);
    \coordinate (2ADV2) at ($(2ADV) + (270:0.5)$);
    \coordinate (2ADV3) at ($(2ADV) + (90:0.25)$);

    \coordinate (3ADV) at  ($(2ADV)+(180:2)$);
    \coordinate (3ADV2) at ($(2ADV2)+(180:2)$);
    \coordinate (3ADV3) at ($(2ADV3)+(180:2)$);

    \coordinate (4ADV) at  ($(2ADV)+(0:2)$);
    \coordinate (4ADV2) at ($(2ADV2)+(0:2)$);
    \coordinate (4ADV3) at ($(2ADV3)+(0:2)$);

    \node[draw=blue!10, rectangle, rounded corners=2.5mm, minimum width=2cm, minimum height=2cm, anchor=center, fill=blue!10](2b2) at (3ADV){};
    \node[scale=0.8, align=center](2b3) at (3ADV2){Player $1$};
    \node[businessman, shirt=blue, scale=1.5](2b1) at (3ADV3){};

    \node[draw=blue!10, rectangle, rounded corners=2.5mm, minimum width=2cm, minimum height=2cm, anchor=center, fill=blue!10](2b2) at (4ADV){};
    \node[scale=0.8, align=center](2b3) at (4ADV2){Player $N$};
    \node[businessman, shirt=blue, scale=1.5](2b1) at (4ADV3){};

    \coordinate (cc1) at  ($(2ADV)+(0:0.5)$);
    \coordinate (cc2) at  ($(2ADV)+(180:0.5)$);
    \fill (2ADV) circle (2pt);
    \fill (cc1)  circle (2pt);
    \fill (cc2)  circle (2pt);

    \coordinate (cc3) at  ($(B21)!0.5!(B3)+(90:0.5)$);
    \coordinate (cc4) at  ($(cc3)+(0:0.5)$);
    \coordinate (cc5) at  ($(cc3)+(180:0.5)$);
    \fill (cc3) circle (1pt);
    \fill (cc4)  circle (1pt);
    \fill (cc5)  circle (1pt);


    \coordinate (s1d) at ($(A11)!0.5!(B11)$);
    \coordinate (s1u) at ($(C1)!0.5!(D1)$);

    \coordinate (s2d) at ($(A21)!0.5!(B21)$);
    \coordinate (s2u) at ($(C2)!0.5!(D2)$);

    \coordinate (s3d) at ($(A31)!0.5!(B31)$);
    \coordinate (s3u) at ($(C3)!0.5!(D3)$);

    \coordinate (p1p) at ($(3ADV)+(270:1)$);
    \coordinate (p2p) at ($(4ADV)+(270:1)$);

    \coordinate (ap)  at ($(ADV) + (90:1)$);
    
    \draw[-{Triangle[length=5pt, width=4.5pt]},line width=0.8pt](p1p)--(s1u);
    \draw[-{Triangle[length=5pt, width=4.5pt]},line width=0.8pt](p1p)--(s2u);
    \draw[-{Triangle[length=5pt, width=4.5pt]},line width=0.8pt](p1p)--(s3u);

    \draw[-{Triangle[length=5pt, width=4.5pt]},line width=0.8pt](p2p)--(s1u);
    \draw[-{Triangle[length=5pt, width=4.5pt]},line width=0.8pt](p2p)--(s2u);
    \draw[-{Triangle[length=5pt, width=4.5pt]},line width=0.8pt](p2p)--(s3u);  

    \draw[-{Triangle[length=5pt, width=4.5pt]},line width=0.8pt] (s1d)--(ap);
    \draw[-{Triangle[length=5pt, width=4.5pt]},line width=0.8pt](s2d)--(ap);
    \draw[-{Triangle[length=5pt, width=4.5pt]},line width=0.8pt](s3d)--(ap);

    \coordinate (stage1) at ($(B1)!0.5!(A1) + (270:0.25)$);
    \coordinate (stage2) at ($(B2)!0.5!(A2) + (270:0.25)$);
    \coordinate (stage3) at ($(B3)!0.5!(A3) + (270:0.25)$);

    \node[scale=0.8, align=center]() at (stage1){Stage $1$};
    \node[scale=0.8, align=center]() at (stage2){Stage $2$};
    \node[scale=0.8, align=center]() at (stage3){Stage $K$};

    \coordinate (pstage1) at ($(B1)!0.5!(A1) + (90:0.5) $);
    \coordinate (pstage2) at ($(B2)!0.5!(A2) + (90:0.5) $);
    \coordinate (pstage3) at ($(B3)!0.5!(A3) + (90:0.5) $);

    \node[scale=0.8, align=center]() at (pstage1){Potential \\ profit};
    \node[scale=0.8, align=center]() at (pstage2){Potential \\ profit};
    \node[scale=0.8, align=center]() at (pstage3){Potential \\ profit};

    \coordinate (lstage1) at ($(B11)!0.5!(A11) + (90:0.5) $);
    \coordinate (lstage2) at ($(B21)!0.5!(A21) + (90:0.5) $);
    \coordinate (lstage3) at ($(B31)!0.5!(A31) + (90:0.5) $);

    \node[scale=0.55, align=center]() at (lstage1){$\Psi_1(\Phi_{1,\mathbf{x}})$};
    \node[scale=0.55, align=center]() at (lstage2){$\Psi_2(\Phi_{2,\mathbf{x}})$};
    \node[scale=0.55, align=center]() at (lstage3){$\Psi_K(\Phi_{K,\mathbf{x}})$};


   
\end{tikzpicture}
\end{adjustbox}
    \caption{\unboldmath Illustration of the lossy market model with $N$ players and $K$ stages. At stage $k\in\Z_K$, player $i\in\mc I$ chooses a decision vector $x_{i,k}\in\R_+^m$, where $m$ represents the number of distinct allocation categories, and participates in the contest for a certain stage profit through its \textit{participation function} $\phi_{i,k}(x_{i,k})\in\R_+$. At the same time, the stage profit loss $\Psi_k(\Phi_{k,\mathbf{x}})$, determined as a function of the total player participation at stage $k$, i.e., $\Phi_{k,\mathbf{x}}=\sum_{i\in\mc I}\phi_{i,k}(x_{i,k})$, is deduced from a predefined fixed value of the stage's total potential profit $W_k\in\R_+$, which in turn reduces the maximal payoff of the players.} 
\label{fig:img_tac}
\end{figure}

To capture these real-world economic scenarios, we focus on games where profitability is directly linked to the balance between supply and demand, and where greater resource investment leads to higher potential returns. However, much of the existing literature centers on models with the so-called fixed prize structure, meaning that the total potential profit for players is predetermined, irrespective of their individual investments~\cite{LIM2014155, TCTD, KIM20181, KIM201731, 9028855, 10115019, 9993253}. Depending on the setup, the stage value is then either divided among the players~\cite{LIM2014155, TCTD, KIM20181, KIM201731} or awarded entirely to the highest investor~\cite{9028855, 10115019, 9993253}. As a consequence, there is often a need to ensure that player interactions conform to certain fairness criteria. Typically, the nature of the application favors some form of fair proportional allocation~\cite{SHAMS2014129, 10.1145, 7402532}, with the well-established Tullock contest being a frequently adopted modeling approach~\cite{LIM2014155, TCTD, KIM20181, Ewerhart2020, ezra, Warneryd2012}. While it ensures profits are proportional to allotted resources, its nonlinear mechanism can cause small differences in resource expenditure to lead to large profit disparities. Coupled with the assumption of perfect profit division, this can result in disproportionately high rewards for marginally higher investment, which does not reflect the real-world scenarios where high demand requires substantial supply to provide large profits. For example, if two players invest equally small amounts in a stage contest, they would split the profit equally despite minimal total investment. Moreover, if this amount is insignificant for one but substantial for the other, by doubling its investment, the first player could secure twice the profit of the other player, while still using only a small portion of their budget. The situation becomes more unrealistic if one player does not participate at all. Then, the other player can in theory secure the whole, possibly very high profit, with infinitesimal investment. With this in mind, we recognize the need for a modeling framework that accounts for potential profit loss due to insufficient player participation that can also maintain a certain established notion of fairness.  

To align with the aforementioned principles, this paper introduces a general multi-stage market model with lossy stage profits as illustrated in Figure~\ref{fig:img_tac}. Building on our preliminary works~\cite{maljkovic2024blotto} and~\cite{maljkovic2024cdc}, that touched upon these ideas within the context of Blotto and Receding Horizon two-player games, we generalize the proposed setup and allow any number of players to participate in the market. From the structural perspective, we propose a setup that is similar to single-stage, incomplete information frameworks of~\cite{Ewerhart2020, ezra, Warneryd2012} in a sense that players' stage profits are obtained as the difference between the stage market payoff and the cost of participation. However, in~\cite{Ewerhart2020, ezra, Warneryd2012} players compete for a single prize, with the Tullock contest-like nature arising from the assumption that a player's winning probability is directly proportional to exerted efforts, thus eliminating the need to consider any form of collective loss. Conversely, we formalize the notions of players' participation functions, the corresponding stage profit loss, and look at the problem from the perspective of an extended game, where the loss model can be seen as a fictitious player always trying to claim a certain amount of the potential profit, but will get less when the non-fictive players invest more. To maintain consistency, we define the notion of fictitious player's participation function but stress its dependence on the chosen form of the stage profit loss. Furthermore, we demonstrate that a Tullock-like structure for the stage payoff arises as a unique solution when weighted fair proportional resource allocation is implemented in the sense of~\cite{Kelly}. As the resulting stage payoff can enjoy virtually any properties based on the chosen form of the fictitious participations, we focus on a simplified scenario that we refer to as \textit{lossy Tullock games}, where the fictitious participations are exogenously given, potentially time-varying, constants. To the best of our knowledge, this is the first instance where a lossy market framework has been proposed and theoretically analyzed from the perspectives of both centralized and self-interested players, all while taking into account the possibility of having multiple allocation categories at each stage. 

With certain structural assumptions, the proposed framework enables computing common viable strategies for the players, e.g., proportionally fair strategies as defined in~\cite{PoF,NICOSIA2017933}, or Nash equilibria~\cite{Rosen}. We begin by revisiting the centralized solutions in the context of lossy games and then shift our focus entirely to the concept of Nash equilibria. Specifically, to illustrate the applicability of the game framework, we study its general properties and characterize sufficient conditions for the existence and uniqueness of Nash equilibria. Finally, we demonstrate that our framework can be viewed as a generalization of both Receding Horizon and Blotto games, testifying about the practical relevance
and applicability of the proposed model. With this in mind, the main contributions of this paper can be summarized as follows:
\begin{itemize}
    \item We present a new class of lossy multi-stage allocation games  that ensure profit can only be generated from high demand when it is met with sufficient supply. Additionally, we show that the well-known concept of weighted proportional fair resource allocation~\cite{Kelly} gives rise to a unique Tullock-like stage profit model.
    \item In the general context of the lossy market model, which does not necessarily rely on the properties of weighted proportional fair resource allocation at the level of market design, we explore both proportionally fair centralized player strategies and self-interested, no-regret, Nash equilibrium strategies. Our main theoretical findings provide sufficient tests for verifying their existence and uniqueness and offer additional technical insights for the setup of lossy Tullock games. As a result, we also provide an iterative, semi-decentralized method guaranteed to converge to a unique Nash equilibrium.
    \item We outline a series of steps that facilitate the application of the proposed games with lossy Tullock contests in a Receding Horizon fashion and explore their connection to the concept of Blotto games. More precisely, we leverage the simple structure of budget constraints in Blotto games to develop an alternative, four-step, semi-analytical method for computing the unique Nash equilibrium and compare its computational complexity to that of the originally proposed iterative method.
    \item Finally, we demonstrate the effectiveness of the proposed framework in a novel numerical case study in the domain of smart mobility.
\end{itemize}
The paper is outlined as follows: the rest of this section is devoted to introducing some basic notation. In Section~\ref{sec:2}, we introduce the general concept of resource allocation in lossy markets, revisit the standard viable solution strategies for players, and derive the generalized extension of the Tullock contests that arises under weighted fair proportional resource allocation stage market design. Moreover, we formally introduce multi stage lossy games for which we provide a general characterization of the Nash equilibria in Section~\ref{sec:3}. In Section~\ref{sec:4}, we focus on a simplified version of lossy stage Tullock contests and provide a theoretical analysis of the common mathematical properties that guarantee existence and uniqueness of the Nash equilibrium. Sections~\ref{sec:5} and~\ref{sec:6} then present our additional insights into the domains of Receding Horizon and Blotto games, and illustrate the proposed framework in a numerical case study. Finally, Section~\ref{sec:7} concludes the paper and presents some ideas for future work.

\textit{Notation:}  Let $\R_{(+)}$ and $\Z_{(+)}$ denote the sets of (non-negative) real and integer numbers. Let $\mathbf{1}_{m}$ and $\mathbf{I}_m$ be the vector of all ones and a unit matrix of size $m$. For any $T\in\Z_+$, we let $\Z_T=\{0,1,2,...,T-1\}$. If $\mc A$ is a finite set of vectors $x_i$, we let $x \defineas \text{col}(x_{i})_{i\in \mc A}$ be their concatenation. For $x\in\R^n$, we let $\text{diag}(x)\in\R^{n\times n}$ denote a diagonal matrix whose elements on the diagonal correspond to vector $x$. If $\mc A$ is a set of $k$ matrices $A_i\in\R^{m\times n}$, then $\text{blkdiag}\{A_i\}_{i\in\mc A}\in\R^{km\times kn}$ denotes the corresponding block-diagonal matrix and $\text{vstack}(A_i)_{i\in\mc A}\in\R^{km\times n}$ denotes their vertical concatenation. We let $\otimes$ be the outer product of vectors.

\section{Resource Allocation in Lossy Markets}\label{sec:2}

For $K\in\N$, consider a setup in which $N$ players, collected in a set $\mc I=\{1,2,\ldots,N\}$, participate in a market spanning $K$ stages. Each player $i\in\mc I$ engages in stage $k\in\Z_K$ by allocating a personal decision vector $x_{i,k}\in\R_+^m$, where $m\in\N$ represents the count of distinct allocation categories per stage, collected in $\mc C$. Hence, for $j\in\mc C$, we let $x_{i,k}^j\geq0$ be the amount of resources player $i$ has allocated for category $j$ in stage $k$, and we let $\mathbf{x}_i\defineas\text{col}(x_{i,k})_{k\in\mc\Z_K}$ describe player $i$'s decision making in the entire market. Typically, player's resources are constrained, 
so we assume existence of stage-feasible sets $\mc X_{i,k}\subseteq\R^m_+$ such that $\mc X_i\defineas\prod_{k\in\Z_K}\mc X_{i,k}$ and
\begin{equation}\label{eq:1}
    \mc X_i=\big\{\mathbf{x}_i\in\R_+^{Km} \mid \mathbf{g}_i^{\text{iq}}(\mathbf{x}_i)\leq\mathbf{0}_{m_i^{\text{iq}}}\land \mathbf{g}_i^{\text{eq}}(\mathbf{x}_i)=\mathbf{0}_{m_i^{\text{eq}}}\big\}\,,
\end{equation}
where $\mathbf{g}_i^{\text{iq}}:\R^{Km}_+\rightarrow\R^{m_i^{\text{iq}}}$ and $\mathbf{g}_i^{\text{eq}}:\R^{Km}_+\rightarrow\R^{m_i^{\text{eq}}}$ encompass $m_i^{\text{iq}}$ inequality and $m_i^{\text{eq}}$ equality constraints. Note that $\mc X_i$ accommodates both stage-level and coupling constraints between the stages imposed on $x_{i,k}$ for $k\in\Z_K$. For this representation of all feasible resource allocations, we introduce the following standard structural assumption:
\begin{assumption}\label{sas:1}
    For every $i\in\mc I$, the set $\mc X_i$ is compact, convex, and satisfies Slater's constraint qualification.
\end{assumption}
\medskip

Let $\mathbf{x}_{-i}\defineas\text{col}(\mathbf{x}_j)_{j\in\mc I\setminus \{i\}}\in\mc X_{-i}$ be the joint allocation strategy of player $i$'s opponents, where $\mc X_{-i}\defineas\prod_{j\in\mc I\setminus \{i\}}\mc X_j$. Then, for a combination of strategies given by $\mathbf{x}_i\in\mc X_i$ and $\mathbf{x}_{-i}\in\mc X_{-i}$, we assume that player $i$'s total profit is given by
\begin{equation}\label{eq:2}
    \mathbf{u}_{i}(\mathbf{x}_{i},\mathbf{x}_{-i})\defineas \sum_{k\in\Z_K}\mathbf{p}_{i,k}(x_{i,k},\mathbf{x}_{-i,k})-\mathbf{c}_{i,k}(x_{i,k},\mathbf{x}_{-i,k})\,,
\end{equation}
where $\mathbf{x}_{-i,k}\defineas\text{col}(x_{j,k})_{j\in\mc I\setminus \{i\}}$, $\mathbf{p}_{i,k}:\R^{Nm}_+\rightarrow\R_+$ and $\mathbf{c}_{i,k}:\R^{Nm}_+\rightarrow\R_+$. Here, $\mathbf{u}_{i,k}(x_{i,k},\mathbf{x}_{-i,k})\defineas\mathbf{p}_{i,k}(x_{i,k},\mathbf{x}_{-i,k})-\mathbf{c}_{i,k}(x_{i,k},\mathbf{x}_{-i,k})$
represents player $i$'s profit at stage $k$ when allocating $x_{i,k}$, with $\mathbf{p}_{i,k}$ being the \emph{stage payoff} and $\mathbf{c}_{i,k}$ being the \emph{participation cost}.

For the general setup in~\eqref{eq:2}, a commonly explored modeling approach for the stage payoffs is the concept of Tullock contests~\cite{LIM2014155, Ewerhart2020, TCTD}. However, within the framework of~\eqref{eq:2}, modeling the market payoff in the context of Tullock contests would imply that the total potential profit for a given stage, denoted by $W_k\in\R_+$, would be entirely distributed among the players, regardless of their level of market participation, i.e., $W_k=\sum_{i\in\mc I}\mathbf{p}_{i,k}(x_{i,k},\mathbf{x}_{-i,k})$ for all $k \in \Z_K$. In real-world scenarios, where profit is driven by meeting high demand, this assumption becomes increasingly unrealistic as fulfilling such demand generally involves substantial resource expenditure. Therefore, it is essential to consider models that factor in profit loss due to insufficient player participation.   

Formally speaking, for every $k\in\Z_K$ and $i\in\mc I$, let $\phi_{i,k}:\R^{m}_+\rightarrow\R_+$ be the \textit{participation function} of player $i$, $\Phi_{k,\mathbf{x}} = \sum_{j \in \mc I} \phi_{j,k}(x_{j,k})$ be the \textit{total participation} of the players at stage $k$ and, $\Psi_k:\prod_{i\in\mc I}\mc X_{i,k}\rightarrow\R_+$ be the corresponding \textit{stage profit loss}. Then, a market with \textit{lossy} stage profits can be introduced through the following definition.
\begin{definition}[Lossy stage profits]\label{def:lossy_stage}
    For every $k\in\Z_K$, let the total potential profit at stage $k$ be given by $W_k\in\R_+$. Then, the stage market is considered \textit{lossy} if the total potential profit is split between the players and a certain portion is forfeited, i.e., the following relation holds:
    \begin{equation}\label{eq:3}
        W_k=\Psi_k(\Phi_{k,\mathbf{x}})+\sum_{i\in\mc I}\mathbf{p}_{i,k}(x_{i,k},\mathbf{x}_{-i,k})\,,
    \end{equation}
    where $\mathbf{p}_{i,k}(x_{i,k},\mathbf{x}_{-i,k})$ is the the individual player $i$'s stage payoff at stage $k$, as stated in~\eqref{eq:2}.
\end{definition}
\medskip
In this paper, we focus on markets that align with Definition~\ref{def:lossy_stage}. Note that we use the notation $\Psi_k(\Phi_{k,\mathbf{x}})$ to emphasize the dependence of the stage profit loss on the total participation of the players. Nevertheless, $\Psi_k$ formally represents a mapping from the product set of players' feasible sets into a real-valued loss. Hence, to establish regularity of markets with lossy stage profits, while ensuring that players must offer
adequate supply in order to capture a significant market share in high-demand scenarios, we postulate the following assumption. 
\begin{assumption}\label{sas:2}
    For $i\in\mc I$ and $k\in\Z_K$, functions $\phi_{i,k}(x_{i,k})$ and $\mathbf{c}_{i,k}(x_{i,k},\cdot)$ are continuous and twice differentiable on $\mc X_{i,k}$, it holds that $\phi_{i,k}(\mathbf{0}_m)=\mathbf{c}_{i,k}(\mathbf{0}_m,\cdot)=0$, and stage profit loss $\Psi_k(\Phi_{k,\mathbf{x}})$ is a decreasing function of $\Phi_{k,\mathbf{x}}$.
\end{assumption}
\medskip

Apart from tackling~\eqref{eq:2} from the perspective of lossy stage profits, we will also focus on scenarios where stage payoffs align with the principle that greater resource investment results in higher outcomes for the players. To enforce such behavior, extensive literature in the areas of communication networks, transportation, and economics often suggests different proportional allocation mechanisms tailored to the specific type of application~\cite{7402532, laur, 10.1145, Kelly}. 

Interestingly, for $K=1$, $|\mc C|=1$, linear costs $\mathbf{c}_{i}(x_{i},\mathbf{x}_{-i})=\alpha_ix_{i}$, and sets $\mc X_i=\{x_i\in\R_+\mid\alpha_ix_i\geq 0\}$, the structure of the total profit~\eqref{eq:2} recovers a version of the well-known user objective in the problem of charging and rate control for elastic traffic in communication networks~\cite{elastic}, though in a form parameterized by other players' decision vectors. In such systems, ensuring fairness at the network level is critical to prevent blocking user access while maximizing overall throughput of the network. Hence,~\cite{Kelly} and~\cite{elastic} draw a link between network performance and various fairness concepts, motivating us to closely examine specific cases of stage profits given by Definition~\ref{def:lossy_stage}, particularly in the light of \textit{weighted proportional fair resource allocation} as outlined in~\cite{Kelly}.  

Namely, to simultaneously account for the stage profit loss and ensure that the stage profit is fairly distributed according to players' participation, and within the framework of~\cite{Kelly}, we consider an extended setup with a fictive player making up for the loss at each stage. To distinguish it from the real players, we introduce its continuously differentiable \textit{fictitious participation} $\varepsilon_k:\R_+\rightarrow\R_+$ at stage $k\in\Z_K$ and make it a function of real players' total participation $\Phi_{k,\mathbf{x}}$. Consequently, fictive player's participation $\varepsilon_k(\Phi_{k,\mathbf{x}})$ becomes directly linked to the stage profit loss $\Psi_k(\Phi_{k,\mathbf{x}})$, which now acts as its payoff at stage $k$.  Broadly speaking, $\varepsilon_k$ should reflect the nature of $\Psi_k(\Phi_{k,\mathbf{x}})$ and guarantee that when there is a sufficient supply from the real players, the participation of the fictive one is small. In the following definition, we adapt the formal description of weighted proportional fair resource allocation from~\cite{Kelly} to this extended setup.
\begin{definition}\label{def:ppf}
    For a fixed $k\in\Z_K$, a set of players $\mc I$, and a fictive participation $\varepsilon_k(\Phi_{k,\mathbf{x}})$, a set of stage payoffs $\{\mathbf{p}_{i,k}\}_{i\in\mc I}$ conform to the principles of \emph{weighted proportional fair resource allocations} if they solve
    \begin{multline}\label{eq:ppf}
        \{\mathbf{p}_{i,k}\}_{i\in\mc I}=\\\argmax_{s\in\mc S_{N}}\sum_{i=1}^N\phi_{i,k}(x_{i,k})\log(s_i)+\varepsilon_k(\Phi_{k,\mathbf{x}})\log\biggl(W_k-\sum_{i=1}^N s_i\biggr)\,,
    \end{multline}
    where $\mc S_{N}\defineas\{s\in\R_+^{N}\mid \sum^{N}_{i=1}s_i\leq W_k\}$.
\end{definition}
\medskip
With Definition~\ref{def:ppf} in place, we can now present a general form of the weighted proportional fair stage payoffs and stage profit loss, along with a necessary condition on fictive participations $\varepsilon_k(\Phi_{k,\mathbf{x}})$ for  Assumption~\ref{sas:2} to hold.
\begin{proposition}\label{prop:ppf}
    For a fixed $k\in\Z_K$, a fictive participation $\varepsilon_k(\Phi_{k,\mathbf{x}})$, and $i\in\mc I$, let $\{\mathbf{p}_{i,k}\}_{i\in\mc I}$ be the stage payoffs solving the optimization problem~\eqref{eq:ppf} in Definition~\ref{def:ppf}. Then, these payoffs satisfy
    \begin{equation}\label{eq:16}
        \mathbf{p}_{i,k}(x_{i,k},\mathbf{x}_{-i,k})\defineas W_k\frac{\phi_{i,k}(x_{i,k})}{\sum_{j\in\mc I}\phi_{j,k}(x_{j,k})+\varepsilon_k(\Phi_{k,\mathbf{x}})}\,,
    \end{equation}   
    resulting in a stage profit loss of the form
    \begin{equation}\label{eq:17}
        \Psi_k(\Phi_{k,\mathbf{x}})=W_k\frac{\varepsilon_k(\Phi_{k,\mathbf{x}})}{\Phi_{k,\mathbf{x}}+\varepsilon_k(\Phi_{k,\mathbf{x}})}\,, 
    \end{equation}
    which satisfies Assumption~\ref{sas:2} if and only if $x\varepsilon_k'(x)-\varepsilon_k(x)<0$ holds for all $x\in\R_+$.
\end{proposition}
\medskip
\begin{proof}
    Note that every solution of~\eqref{eq:ppf} has to be element-wise positive and $\sum_{i=1}^N s_i<W_k$. Then, by looking at the KKT conditions of the Lagrangian $\mc L=\sum_{i=1}^N\phi_{i,k}(x_{i,k})\log(s_i)+\varepsilon_k(\Phi_{k,\mathbf{x}})\log(W_k-\sum_{i=1}^N s_i)-\rho_{\text{inq}}^1(\sum_{i=1}^{N+1}s_i-W_k)+\sum_{i=1}^{N+1}\rho_{\text{inq},i}^2 s_i$, based on complementary slackness, we get 
    \begin{equation}\label{eq:condi}
        \frac{\phi_{i,k}(x_{i,k})}{s_i}-\frac{\varepsilon_k(\Phi_{k,\mathbf{x}})}{W_k-\sum_{i=1}^N s_i}=0\text{\:for\:}i\in\mc I\,.       
    \end{equation}
    Then, by rearranging we get
    $$\sum_{i=1}^N s_i=\frac{W_k-\sum_{i=1}^N s_i}{\varepsilon_k(\Phi_{k,\mathbf{x}})}\sum_{i=1}^N \phi_{i,k}(x_{i,k}),$$
    which yields $\sum_{i=1}^N s_i=\frac{\Phi_{k,\mathbf{x}}}{\Phi_{k,\mathbf{x}}+\varepsilon_k(\Phi_{k,\mathbf{x}})}$. Then, by plugging this back into~\eqref{eq:condi}, we get that $s_i$ recovers the form given by~\eqref{eq:16} and that the resulting stage profit loss $\Psi_k(\Phi_{k,\mathbf{x}})=W_k-\sum_{i=1}^N s_i$ recovers~\eqref{eq:17}. Finally, for~\eqref{eq:17} to satisfy Assumption~\ref{sas:2}, it has to be a decreasing function of $\Phi_{k,\mathbf{x}}$. This means that its derivative with respect to $\Phi_{k,\mathbf{x}}$ has to be negative, directly yielding $x\varepsilon_k'(x)-\varepsilon_k(x)<0$ for all $x\in\R_+$.
\end{proof}
\medskip
From the structural perspective, the stage payoff~\eqref{eq:16} resembles the standard Tullock contest and under a mild and fairly realistic assumption that $\varepsilon_k(\Phi_{k,\mathbf{x}})$ is non-increasing, the Assumption~\ref{sas:2} directly holds. However, the degree of freedom provided by the functional form of $\varepsilon_k(\Phi_{k,\mathbf{x}})$ allows to explicitly influence the properties of $\mathbf{u}_{i}(\mathbf{x}_i,\mathbf{x}_{-i})$ typically required for the existence of common solution strategies such as system optima, fair solutions or Nash equilibria. Therefore, in the following subsection we focus on viable solution concepts for markets with lossy stage profits.

\subsection{Solution concept for markets with lossy stage profits}\label{subsec:2a}

If the nature of the problem allows a collaborative centralized computation of the players' strategies, the primary focus of the system operator typically lies in computing either a \textit{utilitarian strategy}, i.e., a system optimum that maximizes the sum of players' utilities, or a \textit{fair strategy}, which adheres to specific fairness norms as perceived by the operator.

Namely, let us denote with $\mathbf{x}\defineas\text{col}(\mathbf{x}_j)_{j\in\mc I}\in\mc X$ the joint allocation strategy of all players, with $\mc X\defineas\prod_{j\in\mc I}\mc X_j$. Then, for the general lossy market described by Definition~\ref{def:lossy_stage}, computing the System Optimum (SO) boils down to minimizing combined total profit loss and participation costs, which corresponds to solving the optimization problem:
\begin{equation}\label{eq:xs}
    \bar{\mathbf{x}}^{\text{SO}}\in\argmax_{\mathbf{x}\in\mc X} -\sum_{k\in\Z_K}\Psi_k(\Phi_{k,\mathbf{x}})-\sum_{k\in\Z_k}\sum_{i\in\mc I}\mathbf{c}_{i,k}(x_{i,k},\mathbf{x}_{-i,k})\,.
\end{equation}
It is known that the system optimum may result in inequities among the players, forcing some of them to settle for disproportionate shares of the total $K$-stage market. Thus, a fair mechanism aligned with globally accepted norms is typically required. Under the umbrella of weighted proportional fair resource allocation, we can also establish a relation to player strategies that follow widely-recognized \textit{propotional fairness} (PF) principles~\cite{PoF,NICOSIA2017933}. 
\begin{definition}[PF strategy]\label{def:PF}
    For the market defined by~\eqref{eq:2} and~\eqref{eq:3}, a joint strategy $\bar{\mathbf{x}}^{\text{pf}}\in\mc X$ conforms to PF principles if no other strategy $\mathbf{x}\in\mc X$ results in a total relative improvement for a subset of players that outweighs the total relative loss experienced by the other players, i.e., 
\begin{equation}\label{eq:6}
    \sum_{i\in\mc I}\frac{\mathbf{u}_{i}\bigl(\mathbf{x}_i,\mathbf{x}_{-i}\bigr)-\mathbf{u}_{i}\bigl(\bar{\mathbf{x}}_{i}^{\text{pf}},\bar{\mathbf{x}}_{-i}^{\text{pf}}\bigr)}{\mathbf{u}_{i}\bigl(\bar{\mathbf{x}}_{i}^{\text{pf}},\bar{\mathbf{x}}_{-i}^{\text{pf}}\bigr)}\leq 0\,.
\end{equation}
\end{definition}
\medskip
If such a PF strategy exists for a given set of profit functions, it can be identified by solving an optimization problem similar to the one described in Definition~\ref{def:ppf}, as demonstrated in the following proposition.
\begin{proposition}\label{prop:pf}
    For the market defined by~\eqref{eq:2} and~\eqref{eq:3} such that $\mathbf{u}_i(\mathbf{x}_i,\mathbf{x}_{-i})>0$ for all $i\in\mc I$ and $\mathbf{x}\in\mc X$, a joint strategy $\bar{\mathbf{x}}^{\text{pf}}\in\mc X$ that conforms to PF principles solves the following optimization problem:
    \begin{equation}\label{eq:pfeq}
        \bar{\mathbf{x}}^{\text{pf}}\in\argmax_{\mathbf{x}\in\mc X} \sum_{i\in\mc I}\log(\mathbf{u}_i(\mathbf{x}_i,\mathbf{x}_{-i}))\,.
    \end{equation}
\end{proposition}
\medskip
\begin{proof}
    We start by observing that $\overline{\mathbf{x}}^{\text{pf}}$ has to maximize the product of all players' utility functions $\mathbf{u}_i(\mathbf{x}_i,\mathbf{x}_{-i})$. Indeed, let $\mathbf{x}\in\mc X$ be any feasible joint allocation strategy and let $$\eta_i=\frac{\mathbf{u}_i(\mathbf{x}_i,\mathbf{x}_{-i})}{\mathbf{u}_i(\overline{\mathbf{x}}^{\text{pf}}_i,\overline{\mathbf{x}}^{\text{pf}}_{-i})}$$ be defined for every $i\in\mc I$. Then, rearranging~\eqref{eq:6} directly yields $\sum_{i\in\mc I}\eta_i\leq N$, which combined with the inequality between the arithmetic and the geometric mean gives
    \begin{equation}
        \biggl(\frac{\prod_{i\in\mc I}\mathbf{u}_i(\mathbf{x}_i,\mathbf{x}_{-i})}{\prod_{i\in\mc I}\mathbf{u}_i(\overline{\mathbf{x}}^{\text{pf}}_i,\overline{\mathbf{x}}^{\text{pf}}_{-i})}\biggr)^{\frac{1}{N}}=\biggl(\prod_{i\in\mc I}\eta_i\biggr)^{\frac{1}{N}}\leq\frac{1}{N}\sum_{i\in\mc I}\eta_i\leq 1\,,
    \end{equation}
    which guarantees that $\prod_{i\in\mc I}\mathbf{u}_i(\mathbf{x}_i,\mathbf{x}_{-i})\leq \prod_{i\in\mc I}\mathbf{u}_i(\overline{\mathbf{x}}^{\text{pf}}_i,\overline{\mathbf{x}}^{\text{pf}}_{-i})$. Since, $\log(\cdot)$ is increasing, maximizing the product of players' utilities is equivalent to maximizing $\log(\prod_{i\in\mc I}\mathbf{u}_i(\mathbf{x}_i,\mathbf{x}_{-i}))$, which directly yields the optimization problem~\eqref{eq:pfeq}.
\end{proof}
\medskip
\begin{remark}
    In general, if $\mathbf{u}_i(\mathbf{x}_i,\mathbf{x}_{-i})$ is (strictly) concave in $\mathbf{x}\in\mc X$ for every $i\in\mc I$, then Proposition~\ref{prop:pf} guarantees the existence of a (unique) PF strategy based on~\cite[Th.5.1]{rock}.
\end{remark}
\medskip

We can interpret the optimization problem~\eqref{eq:pfeq} as a modified version of the weighted proportional fair resource allocation~\eqref{eq:ppf}. Here, each player is assigned an equal weight of one, reflecting equal treatment, and the objective is to optimally divide the total profit of the complete $K$-stage game, subject to each player's feasibility constraints. It is important, however, to recognize that Proposition~\ref{prop:ppf} represents a design choice that only guarantees a weighted proportional fair distribution of potential profit at the stage level. Even if the market design follows Definition~\ref{def:ppf}, special attention must be given when computing the joint allocation strategy to ensure that the players' outcomes align with the principles in Definition~\ref{def:PF}.    

On the other hand, when dealing with self-interested parties, finding a viable solution becomes arguably more challenging as centralized computation is typically not an option. In such scenarios, the interactions between the players establish a game, typically characterized by a set of coupled best-response optimization problems. In the following, we proceed to formally define and analyze the concept of \textit{lossy games}.

\subsection{Lossy games}\label{subsec:2b}

For a market model with previously introduced lossy stage profits, we can define the corresponding game played between the players when they act in self-interested manner. 
\begin{definition}[Lossy game]\label{def:lgames}
    Consider a fixed number of stages $K\in\Z_+$ and a set of players $\mc I$. Each player $i \in \mc I$ has its own constraint set $\mc X_{i}$ satisfying Assumption~\ref{sas:1}, as well as a stage-dependent payoff function and participation cost function, such that its profit is given by~\eqref{eq:2}. Additionally, assume that each player has a stage-dependent participation function $\phi_{i,k}$ such that the resulting stage profit is lossy according to Definition~\ref{def:lossy_stage} and the participation cost and function are such that Assumption~\ref{sas:2} holds. Under these assumptions, we define the $K$-stage \emph{lossy game} $\mc G(K,\mc I)$, as a game where each player $i \in \mc I$ selects its allocation strategy according to
    \begin{equation}\label{eq:game}
    \max_{\mathbf{x}_i\in\mc X_i}\sum_{k\in\Z_K}\mathbf{u}_{i,k}(x_{i,k},\mathbf{x}_{-i,k}) \,.
    \end{equation}
    
\end{definition}
\medskip

Depending on how the general notion of stages is interpreted, the aggregate structure of $\mc G(K, \mc I)$ in~\eqref{eq:game} can recover various instances of games found in the literature. Specifically, when the players' constraints focus solely on the resource-splitting aspect, this structure typically aligns with the concept of Blotto games (BG)~\cite{KIM20181, 9993253,10115019, maljkovic2024blotto}. Alternatively, if the constraint sets also include a temporal element, meaning that there is an internal state of each player that evolves from one stage to the next based on the chosen stage participation, then the setup aligns well with the Receding Horizon games (RHG)~\cite{6994293,9992497,9354436,maljkovic2024cdc}. In any case, a common no-regret solution for all players in such games is the concept of a Nash equilibrium, formally introduced in the following definition. 
\begin{definition}[Nash equilibrium]\label{def:1}
    A joint strategy $\overline{\mathbf{x}}\in\mc X$ is a Nash equilibrium (NE) of the lossy game $\mc G$ in Definition~\ref{def:lgames}, if for all $i\in\mc I$ and all $\mathbf{x}_i\in\mc X_i$ it holds that $$\mathbf{u}_i(\overline{\mathbf{x}}_i,\overline{\mathbf{x}}_{-i})\geq\mathbf{u}_i(\mathbf{x}_i,\overline{\mathbf{x}}_{-i}).$$
\end{definition}
\medskip

In what follows, we focus on identifying and computing the Nash equilibria of general lossy games $\mc G$ as in Definition~\ref{def:lgames}. We begin by showing that the standard NE existence result holds trivially under our assumptions. In contrast, verifying uniqueness is generally more involved. To tackle this, we exploit the aggregative lossy structure of the players' profit functions and propose a novel sufficient condition test to guarantee uniqueness of the Nash equilibrium. We then extend our analysis to a specific subclass of lossy games with simplified stage payoffs~\eqref{eq:16}, referred to as \textit{lossy Tullock games}, in which the fictitious participations are fixed to a predefined set of constants $\varepsilon_k>0$. For this subclass, we demonstrate that the Nash equilibrium can be computed in a distributed manner with provable convergence guarantees.

\section{Characterization of Nash Equilibria in Lossy Games}\label{sec:3}

We start by noting that $\bar{\mathbf{x}}\in\mc X$ will be the NE of a lossy game~$\mc G$ given by~\eqref{eq:game}, if and only if, for every player $i\in\mc I$, the strategy $\bar{\mathbf{x}}_i\in\mc X_i$ solves the best-response problem:
\begin{mini}
    {\mathbf{x}_i\in\mc X_i}{-\sum_{k\in\Z_K}\mathbf{u}_{i,k}(x_{i,k},\bar{\mathbf{x}}_{-i,k}) \label{maxi:op1}}
    {}{}
    \addConstraint{\mathbf{g}_i^{\text{iq}}(\mathbf{x}_i)\leq\mathbf{0}_{m_i^{\text{iq}}}\land \mathbf{g}_i^{\text{eq}}(\mathbf{x}_i)=\mathbf{0}_{m_i^{\text{eq}}}.}{}{}
    \end{mini}
When the players' profits are concave, then the KKT conditions of the individual best-response optimization problems~\eqref{maxi:op1} associated with the Langrangian
\begin{equation}
    \mc L_i\bigl(\mathbf{x}_i,\mathbf{x}_{-i},\lambda_i,\nu_i\bigr)\defineas -\mathbf{u}_i(\mathbf{x}_i,\mathbf{x}_{-i})+\lambda_i^T\mathbf{g}_i^{\text{iq}}(\mathbf{x}_i)+\nu_i^T\mathbf{g}_i^{\text{eq}}(\mathbf{x}_i)\,
\end{equation}
provide a standard optimality test that can be used to verify if $\bar{\mathbf{x}}\in\mc X$ is a NE of $\mc G$. 
\begin{lemma}[Optimality test]\label{lema:2}
    Consider a lossy game $\mc G$ given by Definition~\ref{def:lgames}, where the players' profits $\mathbf{u}_i(\mathbf{x}_i,\cdot)$ are concave in $\mathbf{x}_i\in\mc X_i$. For players $i\in\mc I$, let the best-response optimization problem related to~\eqref{eq:game} be defined by~\eqref{maxi:op1}, and for $\overline{\mathbf{x}}\in\mc X$ and $i\in\mc I$, let the sets of active inequality constraints in~\eqref{maxi:op1} be given by $\mc A_i(\overline{\mathbf{x}}_i)\defineas\{j\in\{1,\ldots,m_i^{\text{iq}}\}\:\:|\:\:\mathbf{g}_{i,j}^{\text{iq}}(\overline{\mathbf{x}}_i)=0\},$ where $\mathbf{g}_{i,j}^{\text{iq}}(\mathbf{x}_i)$ is the $j$-th row of $\mathbf{g}_{i}^{\text{iq}}(\mathbf{x}_i)$. Let $\delta_i^*(\overline{\mathbf{x}})\in\R_+$ be the optimal value of the problem
    \begin{mini}
        {\lambda_i,\nu_i}{\norm{\nabla_{\mathbf{x}_i} \mc L_i\bigl(\mathbf{x}_i,\mathbf{x}_{-i},\lambda_i,\nu_i\bigr)\mid_{\mathbf{x}=\overline{\mathbf{x}}}}^2_{2} \label{maxi:op2}}
        {}{}
        \addConstraint{\lambda_i^j\geq 0 \text{ for }j\in\mc A_i(\overline{\mathbf{x}}_i)}{}{}
        \addConstraint{\lambda_i^j=0 \text{ otherwise}.}{}{}
    \end{mini}
    Then, $\overline{\mathbf{x}}\in\mc X$ solves all the best-response optimization problems~\eqref{maxi:op1} for $i\in\mc I$ if and only if $\delta_i^*(\overline{\mathbf{x}})=0$ for all $i\in\mc I$.
\end{lemma}
\medskip
\begin{proof}
    For each player $i\in\mc I$, the condition $\delta_i^*(\overline{\mathbf{x}})=0$ directly implies the existence of feasible $\lambda_i$ and $\nu_i$ such that $-\nabla_{\mathbf{x}_i}\mathbf{u}_i(\overline{\mathbf{x}}_i,\overline{\mathbf{x}}_{-i})+\nabla_{\mathbf{x}_i}(\lambda_i^T\mathbf{g}_i^{\text{iq}}(\overline{\mathbf{x}}_i))+\nabla_{\mathbf{x}_i}(\nu_i^T\mathbf{g}_i^{\text{eq}}(\overline{\mathbf{x}}_i))=0$, which is exactly the stationarity condition of~\eqref{maxi:op1} for $\mathbf{x}_i$. Conversely, the feasibility of $\lambda_i$, as dictated by the constraints of~\eqref{maxi:op2}, corresponds exactly to the complementarity slackness constraint of~\eqref{maxi:op1}. Since the concavity of $\mathbf{u}_i(\mathbf{x}_i,\cdot)$ under the  Assumption~\ref{sas:1} makes~\eqref{maxi:op1} convex, the optimality of $\mathbf{x}_i\in\mc X_i$ is equivalent to finding a solution of the KKT system, thereby completing the proof.
\end{proof}
\medskip
As shown later, Lemma~\ref{lema:2} will prove instrumental in designing stopping criteria for the iterative distributed computation of a NE. Additionally, beyond offering a verification method through a convex program, the concavity of the players' profits inherently gives rise to a  Nash equilibrium of $\mc G$.
\begin{theorem}\label{th:1}
   Let a lossy game $\mc G$ be given by Definition~\ref{def:lgames} and the players' profits $\mathbf{u}_i(\mathbf{x}_i,\cdot)$ be concave in $\mathbf{x}_i\in\mc X_i$. Then $\mc G$ admits a Nash equilibrium. 
\end{theorem}
\medskip
\begin{proof}
    Since $\mathbf{u}_i(\mathbf{x}_i,\mathbf{x}_{-i})$ is continuous in $\mathbf{x}\in\mc X$ and concave in $\mathbf{x}_i\in\mc X_i$, we can invoke~\cite[Th.1]{Rosen} to prove the existence of a NE as Assumption~\ref{sas:1} ensures that $\mc X_i$ is compact, convex and satisfies Slater's constraint qualification.
\end{proof}
\medskip

The existence guarantees are a fairly standard result in the literature, arising directly from the concavity of the profits and the structure of the feasible sets. However, establishing a sufficient condition for the uniqueness of the NE involves a more challenging procedure. We adopt the following simplifying notation related to Jacobian and Hessian matrices:
    \begin{equation}
        \frac{\partial^2\mathbf{u}_r}{\partial\mathbf{p}\partial\mathbf{q}}\defineas\mathbf{D}_{\mathbf{p}}\nabla_{\mathbf{q}}\mathbf{u}_r\:\land\:\frac{\partial^2\mathbf{u}_r}{\partial\mathbf{p}^2}\defineas\mathbf{Hess}_{\mathbf{p}}(\mathbf{u}_r)=\mathbf{D}_{\mathbf{p}}\nabla_{\mathbf{p}}\mathbf{u}_r\,.
    \end{equation}
Then, for players $i\in\mc I\setminus\{1,N\}$, let us partition the vector of all other players' strategies as $\mathbf{x}_{-i}^T=[\mathbf{x}_{-i,\text{L}}^T,\:\mathbf{x}_{-i,\text{R}}^T]$, with 
\begin{align}
    \mathbf{x}_{-i,\text{L}}=\text{col}(\mathbf{x}_j)_{j\in \{1, \ldots, i-1\}}:\land\:\mathbf{x}_{-i,\text{R}}=\text{col}(\mathbf{x}_j)_{j\in \{i+1, \ldots, N\}}.
\end{align}
Moreover, let $\mathbf{H}^i_{\mathbf{x}}\in\R^{NKm\times NKm}$ given by
\begin{equation}\label{eq:7}
    \mathbf{H}^i_{\mathbf{x}}=\left[\begin{array}{ccc}
       \frac{\partial^2\mathbf{u}_i}{\partial\mathbf{x}_{-i,L}\partial\mathbf{x}_{-i,L}}  & \mathbf{0}  & \frac{\partial^2\mathbf{u}_i}{\partial\mathbf{x}_{-i,R}\partial\mathbf{x}_{-i,L}}\\
       \mathbf{0}  & \mathbf{0}  & \mathbf{0}\\
        \frac{\partial^2\mathbf{u}_i}{\partial\mathbf{x}_{-i,L}\partial\mathbf{x}_{-i,R}} & \mathbf{0}  & \frac{\partial^2\mathbf{u}_i}{\partial\mathbf{x}_{-i,R}\partial\mathbf{x}_{-i,R}}
    \end{array}\right]\,
\end{equation}
for $i\in\mc I\setminus\{1,N\}$, and given by
\begin{equation}\label{eq:8}
    \mathbf{H}^1_{\mathbf{x}}=\left[\begin{array}{cc}
         \mathbf{0} &  \mathbf{0}\\
         \mathbf{0} & \frac{\partial^2\mathbf{u}_1}{\partial\mathbf{x}_{-1}\partial\mathbf{x}_{-1}}
    \end{array}\right],\:\:\mathbf{H}^N_{\mathbf{x}}=\left[\begin{array}{cc}
         \frac{\partial^2\mathbf{u}_N}{\partial\mathbf{x}_{-N}\partial\mathbf{x}_{-N}} &  \mathbf{0}\\
         \mathbf{0} & \mathbf{0}
    \end{array}\right]
\end{equation}
for $i\in\{1,N\}$, represent the \textit{extended} Hessians of $\mathbf{u}_i(\cdot,\mathbf{x}_{-i})$ evaluated at $\mathbf{x}\in\mc X$, with dimensions of the zero matrices chosen to respect the square nature of $\mathbf{H}^i_{\mathbf{x}}$. Similarly, if we let $\mathbf{M}_{\mathbf{x}}\in\R^{NKm\times NKm}$ be given by
\begin{equation}\label{eq:9}
    \mathbf{M}_{\mathbf{x}}=\text{blkdiag}\biggl\{\text{blkdiag}\biggl\{\frac{\partial^2\mathbf{u}_{i,k}}{\partial x_{i,k}^2}\biggr\}_{k\in\Z_K}\biggr\}_{i\in\mc I}\,,
\end{equation}
then the following theorem provides a sufficient condition test for establishing uniqueness of the Nash equilibrium of $\mc G$.
\begin{theorem}[Uniqueness test]\label{th:2}
     Let a lossy game $\mc G$ be given by Definition~\ref{def:lgames} and the players' profits $\mathbf{u}_i(\mathbf{x}_i,\cdot)$ be concave in $\mathbf{x}_i\in\mc X_i$. Furthermore, for every $p,q\in\mc I$, let 
    \begin{equation}\label{eq:10}
        \frac{\partial^2\mathbf{u}_p}{\partial\mathbf{x}_p\partial\mathbf{x}_q}=\frac{\partial^2\mathbf{u}_p}{\partial\mathbf{x}_q\partial\mathbf{x}_p}
    \end{equation}
    hold and let extended Hessians $\mathbf{H}^i_{\mathbf{x}}$ satisfy~\eqref{eq:7} and~\eqref{eq:8}. Then, $\mc G$ has a unique NE if for any $\mathbf{x}\in\mc X$, it holds that
    \begin{equation}\label{eq:11}
        \mathbf{M}_{\mathbf{x}}-\sum_{i\in\mc I}\mathbf{H}_{\mathbf{x}}^i-\mathbf{Hess}_\mathbf{x}\biggl(\mathbf{C}_{\mathbf{x}}+\sum_{k\in\Z_K}\Psi_k\biggr)\prec 0\,,
    \end{equation}
    where $\mathbf{C}_{\mathbf{x}}=\sum_{i\in\mc I}\sum_{k\in\Z_K}\mathbf{c}_{i,k}(x_{i,k},\mathbf{x}_{-i,k})$.
\end{theorem}
\medskip
\begin{proof}
    Based on~\cite[Th.2]{Rosen}, for $\mc G$ to have a unique NE, it suffices that $\mathbf{G}_{\mathbf{x}}+\mathbf{G}^T_{\mathbf{x}}$ be negative definite for every $\mathbf{x}\in\mc X$, where $\mathbf{G}_{\mathbf{x}}$ is given by 
    \begin{equation}\label{eq:12}
       \mathbf{G}_{\mathbf{x}}\defineas\left[\begin{array}{cccc}
           \frac{\partial^2\mathbf{u}_1}{\partial\mathbf{x}_1^2} & \frac{\partial^2\mathbf{u}_1}{\partial\mathbf{x}_2\partial\mathbf{x}_1} & \cdots & \frac{\partial^2\mathbf{u}_1}{\partial\mathbf{x}_N\partial\mathbf{x}_1}\\
           \frac{\partial^2\mathbf{u}_2}{\partial\mathbf{x}_1\partial\mathbf{x}_2} & \frac{\partial^2\mathbf{u}_2}{\partial\mathbf{x}_2^2} & \cdots & \frac{\partial^2\mathbf{u}_2}{\partial\mathbf{x}_N\partial\mathbf{x}_2}\\
            \hdots & \hdots & \ddots & \hdots\\
           \frac{\partial^2\mathbf{u}_N}{\partial\mathbf{x}_1\partial\mathbf{x}_N} & \frac{\partial^2\mathbf{u}_N}{\partial\mathbf{x}_2\partial\mathbf{x}_N} & \cdots & \frac{\partial^2\mathbf{u}_N}{\partial\mathbf{x}_N^2}
       \end{array}\right]\,.
    \end{equation}
    The shape of~\eqref{eq:2} gives $\nabla_{\mathbf{x}_i}\mathbf{u}_i=\text{col}(\nabla_{x_{i,k}}\mathbf{u}_{i,k})_{k\in\Z_K}$, hence 
    \begin{equation}\label{eq:13}
        \frac{\partial^2\mathbf{u}_i}{\partial\mathbf{x}_i^2}=\text{blkdiag}\biggl\{\frac{\partial^2\mathbf{u}_{i,k}}{\partial x_{i,k}^2}\biggr\}_{k\in\Z_K}\,, 
    \end{equation}
    which yields $\mathbf{M}_{\mathbf{x}}=\text{blkdiag}\bigl\{\frac{\partial^2\mathbf{u}_i}{\partial\mathbf{x}_i^2}\bigr\}_{i\in\mc I}$.
    Observe now that the sum of extended hessians $\sum_{i\in\mc I}\mathbf{H}_{\mathbf{x}}^i$ gives the following matrix
    
    \begin{equation}\label{eq:14}
        \left[\begin{array}{cccc}
           \sum\limits_{j\neq 1}\frac{\partial^2\mathbf{u}_j}{\partial\mathbf{x}_1^2} & \sum\limits_{j\neq 2,1}\frac{\partial^2\mathbf{u}_j}{\partial\mathbf{x}_2\partial\mathbf{x}_1} & \cdots & \sum\limits_{j\neq N,1}\frac{\partial^2\mathbf{u}_j}{\partial\mathbf{x}_N\partial\mathbf{x}_1}\\
           \sum\limits_{j\neq 1,2}\frac{\partial^2\mathbf{u}_j}{\partial\mathbf{x}_1\partial\mathbf{x}_2} & \sum\limits_{j\neq 2}\frac{\partial^2\mathbf{u}_j}{\partial\mathbf{x}_2^2} & \cdots & \sum\limits_{j\neq N,2}\frac{\partial^2\mathbf{u}_j}{\partial\mathbf{x}_N\partial\mathbf{x}_2}\\
            \hdots & \hdots & \ddots & \hdots\\
           \sum\limits_{j\neq 1,N}\frac{\partial^2\mathbf{u}_j}{\partial\mathbf{x}_1\partial\mathbf{x}_N} & \sum\limits_{j\neq 2,N}\frac{\partial^2\mathbf{u}_j}{\partial\mathbf{x}_2\partial\mathbf{x}_N} & \cdots & \sum\limits_{j\neq N}\frac{\partial^2\mathbf{u}_j}{\partial\mathbf{x}_N^2}
       \end{array}\right]\,
    \end{equation}
    and that for any $\mc A\subset\mc I$ and $p,q\in\mc A$ it holds that
    \begin{align}
        &\sum\limits_{j\notin\mc A}\frac{\partial^2\mathbf{u}_j}{\partial\mathbf{x}_p\partial\mathbf{x}_q}=\frac{\partial^2}{\partial\mathbf{x}_p\partial\mathbf{x}_q}\biggl[\sum\limits_{i\in\mc I}\mathbf{u}_i-\sum\limits_{j\in\mc A}\mathbf{u}_j\biggr]=\\\label{eq:15}
        &=-\frac{\partial^2}{\partial\mathbf{x}_p\partial\mathbf{x}_q}\biggl[\sum\limits_{k\in\Z_K}\Psi_k+\mathbf{C}_{\mathbf{x}}\biggr]-\sum\limits_{j\in\mc A}\frac{\partial^2\mathbf{u}_j}{\partial\mathbf{x}_p\partial\mathbf{x}_q}\,.
    \end{align}
    By plugging in~\eqref{eq:15} into~\eqref{eq:14} entry-wise, and combining it with condition~\eqref{eq:10} and definition~\eqref{eq:9}, we finally get 
    \begin{equation}        \mathbf{G}_{\mathbf{x}}+\mathbf{G}_{\mathbf{x}}^T=\mathbf{M}_{\mathbf{x}}-\sum_{i\in\mc I}\mathbf{H}_{\mathbf{x}}^i-\mathbf{Hess}_\mathbf{x}\biggl(\mathbf{C}_{\mathbf{x}}+\sum_{k\in\Z_K}\Psi_k\biggr)\,,
    \end{equation}
    which facilitates directly applying~\cite[Th.2]{Rosen} and hence completes the proof. 
 \end{proof}
\medskip
\begin{remark}\label{rem:2}
    If $\Phi_{k,\mathbf{x}}$ is concave in $\mathbf{x}$, then under Assumption~\ref{sas:2}, Theorem~\cite[Th.5.1]{rock} ensures that $\mathbf{x}\mapsto
    \Psi_k$ is convex for every $k\in\Z_K$, yielding $\mathbf{Hess}_{\mathbf{x}}\bigl(\sum_{k\in\Z_K}\Psi_k\bigr)\succeq 0$.
\end{remark}\medskip
For many games, the uniqueness test can be easily verified by examining the convexity of different parts of the players' profit functions with respect to personal decision vectors. Firstly, the definiteness of $\mathbf{M}_{\mathbf{x}}$ carries information about the concavity of the profit functions. Secondly, the extended Hessians indicate how the decisions of other players affect the personal best-response optimization problem. Given that 
\begin{equation}\label{eq:posdef}
    \mathbf{x}^T\mathbf{H}^i_{\mathbf{x}}\mathbf{x}=\mathbf{x}_{-i}^T\bigl(\mathbf{Hess}_{\mathbf{x}_{-i}}\mathbf{u}_i(\cdot,\mathbf{x}_{-i})\bigr)\mathbf{x}_{-i}\,,
\end{equation}
the extended Hessian $\mathbf{H}^i_{\mathbf{x}}$ preserves the definiteness properties of the corresponding $\mathbf{Hess}_{\mathbf{x}_{-i}}\mathbf{u}_i(\cdot,\mathbf{x}_{-i})$, which can be verified by examining the convexity of the profit functions with respect to vector $\mathbf{x}_{-i}$. As a result, Theorem~\ref{th:2} can be directly used as a tool to design games that inherently satisfy conditions for a unique NE, making it suitable for various practical applications requiring distributed computation.   

Motivated by previous case studies in the domain of traffic signal control~\cite{7402532}, ride-hailing operational management~\cite{maljkovic2024blotto,maljkovic2024cdc}, communication~\cite{4604747}, auctions~\cite{bravo}, bandwidth allocation~\cite{laur}, etc., the following section will introduce a natural extension of Tullock contests that takes into account the loss of market share at each stage.

\section{Games with Lossy Tullock contests}\label{sec:4}

As mentioned in Section~\ref{sec:2}, for the remaining part of this work we will focus on markets with a modified version of the standard Tullock contest, where an exogenously given set of constant parameters $\varepsilon_k>0$ for $k\in\Z_K$, gathered in the set $\mc E$, dictates the incurred loss $\Psi_k(\Phi_{k,\mathbf{x}})$. Formally speaking, we assume a parameterized version of the lossy game in Definition~\ref{def:lgames} given by~\eqref{eq:game}, i.e., $\mc G(K,\mc I 
 \:|\: \mc E)$, that admits market payoffs adhering to proportional fair resource allocation as in~\eqref{eq:16}, with $\varepsilon_k(\Phi_{k,\mathbf{x}})=\varepsilon_k$ for every $k\in\Z_K$. 
 \begin{definition}[Lossy Tullock games]\label{def:lssTg}
    Consider a lossy game $\mc G$ given in Definition~\ref{def:lgames}, along with a set of positive constants $\mc E=\{\varepsilon_k\}_{k\in\Z_K}$. Suppose that the stage payoffs are given by
    \begin{equation}\label{eq:ltprofit}
        \mathbf{p}_{i,k}(x_{i,k},\mathbf{x}_{-i,k})\defineas W_k\frac{\phi_{i,k}(x_{i,k})}{\sum_{j\in\mc I}\phi_{j,k}(x_{j,k})+\varepsilon_k}\,,
    \end{equation}
    for each player $\in\mc I$ and stage $k\in\Z_k$. We refer to this game as the \emph{lossy Tullock game} and denote it by $\mc G(K,\mc I\:|\:\mc E)$.    
 \end{definition}
\medskip
 
 Although simple in nature, the choice of constant fictitious participations directly implies that by assigning a high value to~$\varepsilon_k$, we can ensure that players must offer adequate supply in order to capture a significant market share when the demand is high. Additionally, contests of this general form can exhibit various favourable properties when participation functions and costs are properly chosen. Namely, it can easily be verified that~\eqref{eq:ltprofit} is concave in $\phi_{i,k}(\cdot)$. Therefore, we propose a lemma ensuring the overall concavity of $\mathbf{u}_i(\mathbf{x}_i,\mathbf{x}_{-i})$ with respect to~$\mathbf{x}_i$ for different players' participation functions. 
\begin{lemma}[Concavity]\label{lema:3}
    Let $\mathbf{u}_i(\mathbf{x}_i,\mathbf{x}_{-i})$ be the total profit of player $i\in\mc I$ participating in a lossy Tullock game in Definition~\ref{def:lssTg}. Furthermore, let us define
    \begin{equation}\label{eq:18}
        \mathbf{f}_{1,k}^i\defineas\frac{W_k(\sum_{j\in\mc I\setminus \{i\}}\phi_{j,k}(x_{j,k})+\varepsilon_k)}{\bigl(\Phi_{k,\mathbf{x}}+\varepsilon_k\bigr)^2}\:\land\:\mathbf{f}_{2,k}^i\defineas\frac{-2\mathbf{f}_{1,k}^i}{\Phi_{k,\mathbf{x}}+\varepsilon_k} \,.
    \end{equation}  
    Then, $\mathbf{u}_i(\mathbf{x}_i,\mathbf{x}_{-i})$ is concave in $\mathbf{x}_i$ if and only if
    \begin{equation}\label{eq:19}
    \mathbf{f}_{1,k}^i\frac{\partial^2\phi_{i,k}}{\partial x_{i,k}^2}+\mathbf{f}_{2,k}^i\nabla_{x_{i,k}}\phi_{i,k}\otimes\nabla_{x_{i,k}}\phi_{i,k}-\frac{\partial^2\mathbf{c}_{i,k}}{\partial x_{i,k}^2}\preceq 0
    \end{equation}
    holds for $k\in\Z_K$.
\end{lemma}
\medskip
\begin{proof}
    By observing that $\nabla_{\mathbf{x}_i}\mathbf{u}_i=\text{col}(\nabla_{x_{i,k}}\mathbf{u}_{i,k})_{k\in\Z_K}$, and with a slight abuse of notation, we directly have
    \begin{equation}\label{eq:20}
        \nabla_{\mathbf{x}_i}\mathbf{u}_i=\text{col}\biggl(\mathbf{f}_{1,k}^i\frac{\partial\phi_{i,k}}{\partial x_{i,k}}-\frac{\partial\mathbf{c}_{i,k}}{\partial x_{i,k}}\biggr)_{k\in\Z_K}\,.
    \end{equation}
    Recall that for a scalar and vector valued functions $f(\mathbf{x})$ and $\mathbf{g}(\mathbf{x})$, the following identity holds
    \begin{equation}\label{eq:21}
        \mathbf{D}_{\mathbf{x}}(f(\mathbf{x})\mathbf{g}(\mathbf{x}))=\mathbf{g}(\mathbf{x})\otimes\nabla_{\mathbf{x}}f(\mathbf{x})+f(\mathbf{x})\mathbf{D}_{\mathbf{x}}\mathbf{g}(\mathbf{x})\,.
    \end{equation}
    By direct calculation we have $\nabla_{x_{i,k}}\mathbf{f}_{1,k}^i=\mathbf{f}_{2,k}^i\frac{\partial\phi_{i,k}}{\partial x_{i,k}}$, hence
    $$\frac{\partial^2\mathbf{u}_i}{\partial\mathbf{x}_i^2}=\text{blkdiag}\{\Delta_k\}_{k\in\Z_K},$$ with $\Delta_k\defineas\mathbf{f}_{1,k}^i\frac{\partial^2\phi_{i,k}}{\partial x_{i,k}^2}+\mathbf{f}_{2,k}^i\nabla_{x_{i,k}}\phi_{i,k}\otimes\nabla_{x_{i,k}}\phi_{i,k}-\frac{\partial^2\mathbf{c}_{i,k}}{\partial x_{i,k}^2}$. It is now clear that $\mathbf{u}_i(\mathbf{x}_i,\mathbf{x}_{-i})$ is concave if and only if $\Delta_k\preceq 0$ for every $k\in\Z_K$, which completes the proof. 
\end{proof}
\medskip
Observe that $\mathbf{f}_{1,k}^i>0$, $\mathbf{f}_{2,k}^i<0$ and $\nabla_{x_{i,k}}\phi_{i,k}\otimes\nabla_{x_{i,k}}\phi_{i,k}\succeq 0$ for any $\mathbf{x}\in\mc X$. Hence, for a wide range of configurations, it suffices just to look at the properties of $\phi_{i,k}$ and $\mathbf{c}_{i,k}$ to verify concavity of $\mathbf{u}_{i}$, e.g., for setups with affine player participations and convex participation costs, it is clear that $\mathbf{u}_i(\mathbf{x}_i,\mathbf{x}_{-i})$ remains concave.

Similarly, it is easy to verify that for lossy Tullock contests given by~\eqref{eq:ltprofit} it holds that 
\begin{equation}
            \frac{\partial^2\mathbf{p}_{i,k}}{\partial \phi_{i,k}\partial\phi_{j,k}}=\frac{\partial^2\mathbf{p}_{i,k}}{\partial \phi_{j,k}\partial\phi_{i,k}}\,.
\end{equation}
Therefore, we extend the analysis to facilitate the examination of condition~\eqref{eq:10} for different participation functions and costs, which is necessary for applying the proposed uniqueness test.
\begin{lemma}[Commutativity]\label{lema:4}
   Let $\mathbf{u}_i$ denote the total profit of player $i\in\mc I$ participating in a lossy Tullock game in Definition~\ref{def:lssTg}. Then,
    \begin{equation}\label{eq:22}
        \frac{\partial^2\mathbf{u}_p}{\partial\mathbf{x}_p\partial\mathbf{x}_q}=\frac{\partial^2\mathbf{u}_p}{\partial\mathbf{x}_q\partial\mathbf{x}_p}
    \end{equation}
    holds for all $p,q\in \mc I$, if and only if, 
    \begin{align}\label{eq:23}
        \mathbf{f}_{3,k}^p\frac{\partial\phi_{q,k}}{\partial x_{q,k}}&\otimes\frac{\partial\phi_{p,k}}{\partial x_{p,k}}-\frac{\partial^2 \mathbf{c}_{p,k}}{\partial x_{p,k}\partial x_{q,k}} = \\
        &=\mathbf{f}_{3,k}^p\frac{\partial\phi_{p,k}}{\partial x_{p,k}}\otimes\frac{\partial\phi_{q,k}}{\partial x_{q,k}}-\frac{\partial^2 \mathbf{c}_{p,k}}{\partial x_{q,k}\partial x_{p,k}}
    \end{align}
    holds for all $k\in\Z_K$, where $\mathbf{f}_{3,k}^p\defineas -W_k\frac{\Phi_{k,\mathbf{x}}+\varepsilon_k-2\phi_{p,k}}{(\Phi_{k,\mathbf{x}}+\varepsilon_k)^3}$.
\end{lemma}
\medskip
\begin{proof}
    We start by recalling~\eqref{eq:20} and apply~\eqref{eq:21} for $f=\mathbf{f}_{1,k}^p$ and $\mathbf{g}=\frac{\partial\phi_{p,k}}{\partial x_{p,k}}$, which directly yields
    \begin{equation}
        \frac{\partial^2\mathbf{u}_p}{\partial\mathbf{x}_q\partial\mathbf{x}_p}=\text{blkdiag}\biggl\{\mathbf{f}_{3,k}^p\frac{\partial\phi_{p,k}}{\partial x_{p,k}}\otimes\frac{\partial\phi_{q,k}}{\partial x_{q,k}}-\frac{\partial^2 \mathbf{c}_{p,k}}{\partial x_{q,k}\partial x_{p,k}}\biggr\}_{k\in\Z_K}
    \end{equation}
    since $\mathbf{D}_{\mathbf{x}_q}\frac{\partial\phi_{p,k}}{\partial x_{p,k}}=\mathbf{0}$. On the other hand, we have
    \begin{equation}
        \nabla_{\mathbf{x}_q}\mathbf{u}_p=\text{col}\biggl(-\frac{W_k\phi_{p,k}(x_{p,k})}{(\Phi_{k,\mathbf{x}}+\varepsilon_k)^2}\frac{\partial\phi_{q,k}}{\partial x_{q,k}}-\frac{\partial \mathbf{c}_{p,k}}{\partial x_{q,k}}\biggr)_{k\in\Z_K}\,,
    \end{equation}
    so applying~\eqref{eq:21} for $f=-\frac{W_k\phi_{p,k}(x_{p,k})}{(\Phi_{k,\mathbf{x}}+\varepsilon_k)^2}$ and $\mathbf{g}=\frac{\partial\phi_{q,k}}{\partial x_{q,k}}$, gives
    \begin{equation}
        \frac{\partial^2\mathbf{u}_p}{\partial\mathbf{x}_p\partial\mathbf{x}_q}=\text{blkdiag}\biggl\{\mathbf{f}_{3,k}^p\frac{\partial\phi_{q,k}}{\partial x_{q,k}}\otimes\frac{\partial\phi_{p,k}}{\partial x_{p,k}}-\frac{\partial^2 \mathbf{c}_{p,k}}{\partial x_{p,k}\partial x_{q,k}}\biggr\}_{k\in\Z_K}\,,
    \end{equation}
    which directly completes the proof. 
\end{proof}
\medskip
The condition~\eqref{eq:23} provides a general test which, in certain practical cases, reduces to only verifying the commutativity of the participation costs $\mathbf{c}_{p,k}$. Namely, if the derivatives of the players' participation functions are linearly dependent for every $p,q\in\mc I$, then the first terms on both sides of \eqref{eq:23} will be identical, e.g., if the players use the same affine participation functions and linear participation costs independent of other players' decisions. Moreover, in many practical applications~\cite{9992497,6747393,9732452}, the participation cost of players can be governed by the notion of dynamic prices that depend on the total demand of participants. In the context of multi-stage games, this would correspond to stage participation costs in the form of $\mathbf{c}_{i,k}(x_{i,k},\mathbf{x}_{-i,k})=x_{i,k}^T\sigma_k$, where $\sigma_k\in\R^m$ represents the price vector given by $\sigma_k\defineas \alpha_{k}\sum_{j\in\mc I}x_{j,k}+r_{k}$, and parameterized by some $\alpha_k>0$ and $r_k\in\R^m$. 

With this in mind, we further demonstrate the modelling capabilities of setups with lossy Tullock contests by leveraging Lemmas~\ref{lema:3} and~\ref{lema:4} to construct game instances that inherently give rise to a unique Nash equilibrium.
\begin{proposition}\label{prop:2}
    Let $\mc G(K, \mc I\:|\:\mc E)$ be a lossy Tullock game as in Definition~\ref{def:lssTg}. For $i\in\mc I$, let players' participation functions be given by $\phi_{i,k}(x_{i,k})=w^Tx_{i,k}$, with $w^j\geq 0$ for all $j\in\mc C$ and $\norm{w}>0$. Moreover, let their participation costs be  
    \begin{equation}
        \mathbf{c}_{i,k}(x_{i,k},\mathbf{x}_{-i,k})=x_{i,k}^T\biggl(\alpha_{k}\sum_{j\in\mc I}x_{j,k}+r_{k}\biggr)\,,
    \end{equation}
    with $r_{k}\in\R^m$, and $\alpha_{k}> 0$ being a stage-dependant scaling factor. Then, $\mc G(K, \mc I \:|\: \mc E)$ admits a unique Nash equilibrium.
\end{proposition}
\medskip
\begin{proof}
    To show existence, it is easy to verify that~\eqref{eq:19} reduces to $\mathbf{f}_{2,k}^iww^T-2\alpha_{k}\mathbf{I}_m$. Since $\mathbf{f}_{2,k}^i<0$ and $ww^T\succeq 0$, the players' profits are strictly concave in personal decision vectors. To verify commutativity~\eqref{eq:23}, we observe that $$\frac{\partial\phi_{p,k}}{\partial x_{p,k}}=\frac{\partial\phi_{q,k}}{\partial x_{q,k}}=w \land\frac{\partial^2\mathbf{c}_{p,k}}{\partial x_{p,k}\partial x_{q,k}}=\frac{\partial^2\mathbf{c}_{p,k}}{\partial x_{q,k}\partial x_{p,k}}=\alpha_{k}\mathbf{I} \, .$$ We can now directly apply Theorem~\ref{th:2}. Since players' profits are strictly concave, we have $\mathbf{M}_x\prec 0$. Moreover, the mapping $\mathbf{x}_{-i} \mapsto \sum_{j\neq i}\phi_{j,k}(x_{j,k})+\varepsilon_k$ is concave as its Hessian given by $H_1=\text{blkdiag}\bigl\{\frac{\partial^2\phi_{j,k}}{\partial x_{j,k}^2}\bigr\}_{j\in\mc I\setminus \{i\}}$ is a zero matrix. Hence, given that the mapping $Y\mapsto\frac{\phi_{i,k}(x_{i,k})}{\phi_{i,k}(x_{i,k})+Y}$ is convex and non-increasing for any $\phi_{i,k}(x_{i,k})\geq 0$,~\cite[Th.5.1]{rock} guarantees that $\mathbf{p}_{i,k}$ is convex in $\mathbf{x}_{-i}$. Similarly, $\mathbf{x}_{-i}\mapsto\mathbf{c}_{i,k}$ is concave as its Hessian $H_2=\text{blkdiag}\bigl\{\frac{\partial^2\mathbf{c}_{i,k}}{\partial x^2_{j,k}}\bigr\}_{j\in\mc I\setminus \{i\}}$ is also a zero matrix. Hence, $\mathbf{x}_{-i}\mapsto\mathbf{u}_{i,k}$ is convex for every $k\in\Z_K$, which implies the convexity of $\mathbf{x}_{-i}\mapsto\sum_{k\in\Z_K}\mathbf{u}_{i,k}$, yielding $\mathbf{H}_{\mathbf{x}}^i\succeq 0$ due to~\eqref{eq:posdef}. On the other hand, it is easy to show that $$\sum_{i\in\mc I}\mathbf{c}_{i,k}(x_{i,k},\mathbf{x}_{-i,k})=\overline{x}_k^T\alpha_k\mathbf{I}\overline{x}_k+r_k^T\overline{x}_k\,,$$
    where $\overline{x}_k=\sum_{i\in\mc I}x_{i,k}$. Recall that for a scalar-valued function $f(\mathbf{x})$ and a linear map $\mathbf{g}(\mathbf{x})=A\mathbf{x}$, the chain rule yields
    \begin{equation}\label{eq:30}
        \mathbf{Hess}_{\mathbf{x}}(f\circ\mathbf{g})=A^T\cdot\mathbf{Hess}_{\mathbf{g}}(f)\cdot A\,.
    \end{equation}
    By observing that $\overline{x}_k=A_k\mathbf{x}$ for a properly chosen matrix $A_k$, and applying~\eqref{eq:30} for $f(\mathbf{y})=\mathbf{y}^T\alpha_k\mathbf{I}\mathbf{y}+r_k^T\mathbf{y}$ and $\mathbf{g}(\mathbf{x})=A_k\mathbf{x}$, we directly obtain $\mathbf{Hess}_{\mathbf{x}}(\mathbf{C}_\mathbf{x})\succeq0$ since
    \begin{equation}
        \mathbf{Hess}_{\mathbf{x}}(\mathbf{C}_\mathbf{x})=\sum_{k\in\Z_K}\mathbf{Hess}_{\mathbf{x}}\biggl(\sum_{i\in\mc I}\mathbf{c}_{i,k}\biggr)= \sum_{k\in\Z_K}\alpha_k A_k^TA_k\,,   
    \end{equation}
    and $A_k^TA_k\succeq0$ for every $k\in\Z_K$.
    Finally, since $\mathbf{x}\rightarrow\sum_{i\in\mc I}\phi_{i,k}(x_{i,k})$ is concave, Remark~\ref{rem:2} holds and $\mathbf{Hess}_\mathbf{x}\bigl(\mathbf{C}_{\mathbf{x}}+\sum_{k\in\Z_K}\Psi_k\bigr)\succeq 0$, which completes the proof.
\end{proof}
\medskip

Having established sufficient conditions for existence and uniqueness of the Nash equilibrium, we can now turn our focus to designing a method to compute it. Typically, when the game admits a unique NE by satisfying~\eqref{eq:11}, its computation can be associated with solving a corresponding Variational inequality problem~\cite{VIproblems}. In the next subsection, we revisit a commonly used, semi-decentralized, fixed-point iteration method that leverages the uncoupled nature of the constraint sets $\mc X_i$, and combine it with the Armijo step size rule to design a method with provable convergence to the NE.

\subsection{Distributed computation of the Nash equilibrium}\label{subsec:4a}

As hinted previously, looking at the best-response optimization problems~\eqref{maxi:op1}, it can be observed that the unique NE of a lossy Tullock game $\mc G(K,\mc I \:|\: \mc E)$ coincides with the solution of the Variational Inequality problem $\text{VI}(\mc X,F(\mathbf{x}))$~\cite[Th.2]{Paccagnan2016a}, where $F(\mathbf{x})\defineas\text{col}(-\nabla_{\mathbf{x}_i}\mathbf{u}_{i}(\mathbf{x}_i,\cdot))_{i\in\mc I}$ denotes the game's \textit{pseudo-gradient}. 

Based on the structure of $F(\mathbf{x})$, various iterative methods can be used to compute the solution of $\text{VI}(\mc X,F(\mathbf{x}))$. Therefore, we establish the following properties of $F(\mathbf{x})$ that will allow us to use standard results for fixed-point computation.
\begin{lemma}\label{lema:5}
     Let $\mc G(K, \mc I \:|\: \mc E)$ with the pseudo-gradent $F(\mathbf{x})$ be a lossy Tullock game as in Definition~\ref{def:lssTg}. Moreover, let $\mc G(K, \mc I \:|\: \mc E)$ satisfy the conditions of Theorem~\ref{th:2} and fulfill~\eqref{eq:11}. If for every $i\in\mc I$ and $k\in\Z_K$, $\phi_{i,k}$ and $\mathbf{c}_{i,k}$ are twice differentiable with respect to $x_{i,k}$, then there exist $L,\mu>0$ such that for all $\mathbf{x}_1,\mathbf{x}_2\in\mc X$, it holds that $$\norm{F(\mathbf{x}_1)-F(\mathbf{x}_2)}_2\leq L\norm{\mathbf{x}_1-\mathbf{x}_2}_2,$$
    $$(F(\mathbf{x}_1)-F(\mathbf{x}_2))^T(\mathbf{x}_1-\mathbf{x}_2)\geq\mu\norm{\mathbf{x}_1-\mathbf{x}_2}_2^2\,,$$
    i.e, $F(\mathbf{x})$ is $L$-Lipshitz continuous and $\mu$-strongly monotone. 
\end{lemma}
\medskip
\begin{proof}
    To show that $F(\mathbf{x})$ is Lipschitz continuous, we aim to invoke~\cite[P.2.3.2]{Lips}. For this, it suffices to show that every component of $F(\mathbf{x})$ is Lipschitz continuous. The $l$-th row of $F(\mathbf{x})$ is given by a certain triplet $(i,k,j)\in\mc I\times\Z_K\times\mc C$ and the corresponding map $f_l:\mc X\rightarrow\R$
    \begin{equation}
        f_l(\mathbf{x})\defineas\frac{\partial\mathbf{u}_i(\mathbf{x}_i,\mathbf{x}_{-i})}{\partial x_{i,k}^j}=\mathbf{f}_{1,k}\frac{\partial\phi_{i,k}}{\partial x_{i,k}^j}-\frac{\partial\mathbf{c}_{i,k}}{\partial x_{i,k}^j}\,.
    \end{equation}
    Since $\phi_{i,k}$ and $\mathbf{c}_{i,k}$ are twice differentiable they have to be continuous, rendering both $f_l(\mathbf{x})$ and $\nabla_{\mathbf{x}}f_l$ continuous as well. Since the set $\left\{\norm{\nabla_{\mathbf{x}}f_l(\mathbf{x})} \mid \mathbf{x}\in\mc X\right\}$ is compact due to $\mc X$ being compact, it is clear that $f_l(\mathbf{x})$ is Lipshitz continuous. Finally, strong monotonicity of $F(\mathbf{x})$ follows directly from~\cite[P.2.3.2]{VIproblems} and the fact that $\Gamma=\mathbf{G}_{\mathbf{x}}+\mathbf{G}_{\mathbf{x}}^{T}$, defined by~\eqref{eq:12}, is negative definite due to $\mc G(K, \mc I \:|\: \mc E)$ satisfying~\eqref{eq:11}.
\end{proof}
\medskip

Under the umbrella of conditions on $\mc G(K, \mc I \:|\: \mc E)$ imposed by Lemma~\ref{lema:5}, and for any feasible $\mathbf{x}^0\in\mc X$, the iterative scheme $\mathbf{x}^{(t+1)}=\Pi_{\mc X}\bigl[\mathbf{x}^{(t)}-\gamma F(\mathbf{x}^{(t)})\bigr]$, where $\Pi_{\mc X}(\cdot)$ is the projection operator onto the set $\mc X$, can be executed locally as
    \begin{equation}\label{eq:31}
        \mathbf{x}_i^{(t+1)}=\Pi_{\mc X_i}\bigl[\mathbf{x}_i^{(t)}+\gamma \nabla_{\mathbf{x}_i}\mathbf{u}_i(\mathbf{x}_i^{(t)},\mathbf{x}_{-i}^{(t)})\bigr]\,,
    \end{equation}
and converges to the unique NE of $\mc G(K, \mc I \:|\: \mc E)$ for $0<\gamma<\frac{2\mu}{L^2}$ based on~\cite[T.12.1.2]{VIproblems}. However, this theorem only ensures the existence of a sufficiently small fixed step size $\gamma>0$ that guarantees convergence to the unique NE. It does not provide a way to determine an upper bound on $\gamma$ as the Lipshitz and the strong monotonicity constants are application-specific and depend on the constraint sets $\mc X_i$. In order to circumvent the need to estimate these constants through exhaustive search across the feasible space $\mathcal{X}$ for each game instance, we combine~\eqref{eq:31} with a standard Armijo step-size rule and the optimality test given in Lemma~\ref{lema:2} as outlined in Algorithm~\ref{al:1}.

For a particular combination of a priori chosen parameters $\overline{\gamma}>0$ and $\eta\in(0,1)$, the outer loop of the procedure sets the step size for~\eqref{eq:31} to $\gamma=\eta^l\overline{\gamma}$ at iteration  $l\in\Z_+$.
\begin{algorithm}[tbp]
    \caption{Computing the Nash equilibrium}\label{al:1}
    \begin{algorithmic}[1]
    \State \textbf{Input:}  $K$, $\mc I$, $\mc E$, $\overline{\gamma}$, $\eta$, $\text{tol}$, $t_{\text{out}}$
    \State \textbf{Output:} $\mathbf{x}_i$ for all $i\in\mc I$
    \State $\mc X_i=\text{CreateConstraints}(K)$;\Comment{For $i\in\mc I$}
    \State $\mathbf{x}_i^0=\text{Initialize}(\mc X_i)$;
    \State $l=0;\quad\mathbf{x}\leftarrow\text{col}(\mathbf{x}_i^0)_{i\in\mc I}$;
    \While {$\sum_{i\in\mc I}\mathbf{1}(\delta_i^*(\mathbf{x})<\text{tol})<N$}\Comment{In parallel for $i$}
        \State $\gamma=\eta^l\overline{\gamma};\quad t=0$;
        \State $\mathbf{x}_i^{(t)}\leftarrow\mathbf{x}_i^0$;
        \State $\mathbf{x}_i^{(t+1)}=\Pi_{\mc X_i}\left[\mathbf{x}_i^{(t)}+\gamma \nabla_{\mathbf{x}_i}\mathbf{u}_i(\mathbf{x}_i^{(t)},\mathbf{x}_{-i}^{(t)})\right]$;
        \While{$\norm{\mathbf{x}_i^{(t+1)}-\mathbf{x}_i^{(t)}}>\text{tol}$}
            \State $\mathbf{x}_i^{(t)}\leftarrow\mathbf{x}_i^{(t+1)}$;
            \State $\mathbf{x}_i^{(t+1)}=\Pi_{\mc X_i}\left[\mathbf{x}_i^{(t)}+\gamma \nabla_{\mathbf{x}_i}\mathbf{u}_i(\mathbf{x}_i^{(t)},\mathbf{x}_{-i}^{(t)})\right]$;  
            \State $t\leftarrow t+1$;
            \If{$t>t_{\text{out}}$}
                \State $\textbf{break}$;
            \EndIf    
        \EndWhile
        \State $\mathbf{x}\leftarrow\text{col}(\mathbf{x}_i^{(t+1)})_{i\in\mc I}$
        \State $l\leftarrow l+1$  
     \EndWhile
    \end{algorithmic}
\end{algorithm}
The inner loop is then used to iteratively update the decisions of players until convergence (when the optimality test in Lemma~\ref{lema:2} yields $\delta^*(\mathbf{x})=0$ for every $i\in\mc I$) or timeout (after $t_{\text{out}}$ iterations of the inner loop) are reached. If the conditions of the optimality test given by Lemma~\ref{lema:2} are met, the resulting solution corresponds to the Nash equilibrium of $\mc G(K,\mc I \:|\: \mc E)$. Otherwise, the step size is reduced and the inner-loop is repeated. The entire procedure is guaranteed to terminate after a finite number of steps, as convergence to a NE is ensured for a sufficiently small $\gamma$ according to~\cite[T.12.1.2]{VIproblems}. Finally, note that Algorithm~\ref{al:1} scales well with the number of players due to its distributed nature and the fact that the update step~\eqref{eq:31} requires sharing only the information about the total participation $\Phi_{k,\mathbf{x}}$.

With a general computing procedure in place, the next section will examine the relationship between lossy Tullock contests and specific instances of spatially and temporally distributed competitions. For a class of Blotto games with linear participation costs, we will introduce a semi-analytical, four-step characterization of the unique Nash equilibrium, which can serve as the basis for an alternative computation algorithm. Additionally, we will demonstrate how the concept of lossy Tullock games can be extended to Receding Horizon implementations, allowing us to incorporate internal state dynamics alongside standard budget constraints. 

\section{Connection to Colonel Blotto and Receding Horizon Games}\label{sec:5}

While Blotto and Receding Horizon games originate from different theoretical domains, they share the concept of strategic planning across multiple stages. In Blotto games, resources are allocated spatially across different ``battlefields'', whereas in Receding Horizon games, allocation occurs temporally across different ``time steps''. Both frameworks align well with our notion of stages, yet they differ in the problem formulation and the way the feasible sets $\mc X_i$ are constructed to adhere to the corresponding paradigm. In Blotto games, resources are allocated all at once, and players then compete against each other across various battlefields. On the other hand, Receding Horizon games incorporate the concept of stage ordering, where resources are distributed sequentially while adhering to some underlying system dynamics.
In this section, we will exploit their similarities and illustrate how instances of both of these games can be recovered in the context of proposed games $\mc G(K,\mc I \:|\: \mc E)$ with lossy stage Tullock contests. 

\subsection{Affine Receding Horizon games with lossy Tullock contests}\label{subsec:5a}

In the framework of Receding Horizon games (RHG), each player $i\in\mc I$ aims to optimize the trajectory of its internal state over the horizon of $T\in\N$ future time intervals. At the start of each interval $k\in\Z_T$, players obtain estimates of relevant parameters over the horizon and plan the optimal sequence of actions $u_{i,k}\in\R^{m_{\text{u}}}$ for $k\in\Z_T$. In the specific case of affine RHG, we assume the internal state $y_{i,k}\in\R^{m_{\text{y}}}$ of each player evolves according to a linear state-space model given by
\begin{equation}\label{eq:32}
    y_{i,k+1}=A_iy_{i,k}+B_iu_{i,k}\,,
\end{equation} where $A_i$ and $B_i$ are matrices of appropriate dimensions. To ensure the feasibility of this dynamics, the input vector $u_{i,k}$ is typically subject to stage-level constraints of the form:
\begin{equation}\label{eq:33}
    G_iy_{i,k}+H_iu_{i,k}\leq d_{i,k}\,,
\end{equation}
for $d_{i,k}\in\R^{m_{\text{d}}}$, and $G_i$, $H_i$ of appropriate dimensions.  

To integrate the Receding Horizon framework with the lossy Tullock games in Definition~\ref{def:lssTg}, we adopt a particular form of the stage-level allocation $x_{i,k}\in\R^{m}_+$. In particular, for $m=m_{\text{u}}+1$, we focus on a subclass of lossy Tullock games where 
\begin{equation}
    x_{i,k}^T=\bigl[\varphi_{i,k}(y_{i,k},u_{i,k}), u_{i,k}^T\bigr]\,,
\end{equation}
and $\varphi_{i,k}:\R^{m_{\text{y}}+m_{\text{u}}}\rightarrow\R_+$ is an \emph{auxiliary} affine participation function that depends on the player’s internal state and control input. Specifically, we assume this function is given by
\begin{equation}\label{eq:auxpart}
    \varphi_{i,k}(y_{i,k},u_{i,k})\defineas p_y^Ty_{i,k}+p_u^Tu_{i,k} \,,
\end{equation}
for some vectors $p_y\in\R^{m_{\text{y}}}$ and $p_u\in\R^{m_{\text{u}}}$, which collectively shape the player’s stage-level payoff. Furthermore, if the participation cost depends solely on the stage-level actions $u_{i,k}$, then computing the Nash equilibrium input sequence over the horizon $\Z_T$ effectively boils down to solving the lossy Tullock game $\mc G(K, \mc I \:|\: \mc E)$ in Definition~\ref{def:lssTg} for $K=T$ and linear participation functions of the form $\phi_{i,k}(x_{i,k})=w^T x_{i,k}$, where $w^T=[1,0,\ldots,0]$. 

Therefore, we only have to show that dynamics~\eqref{eq:32}, input constraints~\eqref{eq:33}, and auxiliary participation~\eqref{eq:auxpart} can be cast in the general form of~\eqref{eq:1}. Let $u_{i,k}=S_{u}x_{i,k}$ for properly chosen selection matrix $S_u=[\mathbf{0}\quad\mathbf{I}_{m_{\text{u}}}]$. If we now recall that the state trajectory of the system~\eqref{eq:32} is given by $y_{i,k}=A_i^ky_i^0+\sum_{l=0}^{k-1}A_i^{k-l-1}B_iu_{i,l}$ for $y_{i,0}=y_i^0$, then we can define $z_{i,k}=-A_{i}^ky_i^0$ for every $k\geq 0$, and
\begin{equation}
    \Sigma_{i,k}=\bigl[A_i^{k-1}B_iS_{u}\mid\dots\mid A_i^0B_iS_u\bigr]\,
\end{equation}
for every $k\geq 1$, so that it holds 
\begin{align}\label{eq:34}
    [p_y^T\Sigma_{i,k}\:\mid\:p_u^TS_u-w^T]X_i^{0:k}&=p_y^Tz_{i,k}\\\label{eq:35}
    [G_i\Sigma_{i,k}\:\mid\:H_iS_u]X_i^{0:k}&\leq d_{i,k}+G_iz_{i,k} \,,
\end{align}
with $X_i^{0:k}=\text{col}(x_{i,t})_{t\in\Z_{k+1}}$. Since $\mathbf{x}_i=X_i^{0:T-1}$, we can define $\text{L}_{i,0}^{\text{eq}}=[p_u^TS_u-w^T\:\mid\:\mathbf{0}_{m_{\text{y}}\times m(T-1)}]$, $\text{L}_{i,0}^{\text{inq}}=H_iS_u$, and 
\begin{align}        
 \text{L}_{i,k}^{\text{eq}}&=\bigl[p_y^T\Sigma_{i,k}\:\mid\:p_u^TS_u-w^T\:\mid\:\mathbf{0}_{m_{\text{y}}\times m(T-k-1)}\bigr]\\
\text{L}_{i,k}^{\text{inq}}&=\bigl[G_i\Sigma_{i,k}\:\mid\:H_iS_u\:\mid\:\mathbf{0}_{m_{\text{d}}\times m(T-k-1)}\bigr]
\end{align}
for $k\geq 1$, such that for every $k\geq 0$ we can recast~\eqref{eq:34} and~\eqref{eq:35} as $\text{L}_{i,k}^{\text{eq}}\mathbf{x}_i=p_y^Tz_{i,k}$ and $\text{L}_{i,k}^{\text{inq}}\mathbf{x}_i\leq d_{i,k}+G_iz_{i,k}$, which clearly yields a polytopic constraint set $\mc X_i$ that adheres to~\eqref{eq:1}.

Hence, all the analysis conducted in Sections~\ref{sec:3} and~\ref{sec:4} directly applies, and in order to find optimal planning actions over the horizon, it suffices to solve $\mc G(T,\mc I \:|\: \mc E)$ and extract the segments corresponding to $u_{i,k}$ for each $k\in\Z_T$. It is important to note that from the practical aspect, there are two ways players $i\in\mc I$ could go about controlling their internal state over $T_{\text{total}}\in\N$ time intervals:
\begin{enumerate}
    \item In an open-loop manner, i.e., the exogenous parameters can be well estimated for the whole time frame so the planning horizon can be chosen as $T=T_{\text{total}}$.
    \item In a closed-loop, receding-horizon fashion, i.e., the horizon is $T<T_{\text{total}}$, and at each time step $k$, the players apply only the first element of the control trajectory. 
\end{enumerate}
In the open-loop manner, $T_{\text{total}}$ control inputs are computed only once and directly applied. On the other hand,the receding-horizon implementation requires the complete procedure to be repeated $n_{\text{total}}=T_{\text{total}}-T+1$ times. For time intervals $0\leq k\leq n_{\text{total}}-1$, the planning is executed with horizon $T$ and only the first element of the predicted control trajectory is applied, i.e., the first $T_{\text{total}}-T$ control inputs yield a closed-loop system of the form
\begin{equation}\label{eq:40}
    y_{i,k+1}=A_iy_{i,k}+B_i\kappa_i(y_{i,k},T)\,,
\end{equation}
where $\kappa_i(y_{i,k},T)$ extracts $u_{i,0}$ from the obtained solution $\mathbf{x}_i$ when initialized with $y_i^0=y_{i,k}$. Conversely, the control inputs for the last $T$ time intervals are applied in the open-loop fashion and are calculated in the final execution of the complete procedure, i.e., for $k=n_{\text{total}}$. In Section~\ref{sec:6}, we will demonstrate the performance of both approaches in a numerical case study based on a particular application.

Finally, observe that a spatially distributed, yet structurally simpler, counterpart of Receding Horizon games is embodied by the concept of Blotto games. In the following subsection, we illustrate how the structural properties of Blotto games can be leveraged to design a four-step, closed-form procedure to compute the Nash equilibrium of $\mc G(K, \mc I \:|\: \mc E)$ with linear stage participation costs, for which we can also quantify the execution time given the problem size. 

\subsection{Blotto games with lossy Tullock contests}\label{subsec:5b}

If we adopt certain structural assumptions for the general form of lossy Tullock games, we can recover specific instances that closely align with the well-established literature on Blotto games (BG)~\cite{KIM20181}. Namely, we start by assuming that each player $i\in\mc I$ aims to optimally distribute a fixed budget of $\mathbf{R}_i\in\R_+$ resources among $K$ distinct ``battlefields'' indexed by $k\in\Z_K$, implying that $x_{i,k}\in\R_+$, i.e., $m=1$ and $\mathbf{1}_K^T\mathbf{x}_i=\mathbf{R}_i$. In contrast to RHG, there is no temporal coupling between stages as in~\eqref{eq:32}, meaning that $\mc X_i$ simplifies to $$\mc X_i\defineas\bigl\{\mathbf{x}_i\in\R_+^K\:\mid\:\mathbf{1}_K^T\mathbf{x}_i=\mathbf{R}_i\land x_{i,k}\geq0,\forall k\in\Z_K\bigr\}.$$
From the perspective of players' profits, a BG typically adopts a participation function $\phi_{i,k}(x_{i,k})=x_{i,k}$, so we restrict our focus to lossy Tullock games $\mc G(K, \mc I \:|\: \mc E)$ given by Definition~\ref{def:lssTg}, in which players also face linear stage participation costs, i.e., each player $i \in \mc I$ aims to maximize 
\begin{equation}\label{eq:41}
\max_{\mathbf{x}_i\in\mc X_i}\sum_{k\in\Z_K}\frac{W_kx_{i,k}}{\sum_{j\in\mc I}x_{j,k}+\varepsilon_k}-\beta_kx_{i,k}
\end{equation}
where $\beta_k\in\R$ represents an a priori chosen parameter for every battlefield $k\in\Z_K$. By applying reasoning similar to that in Proposition~\ref{prop:2}, but setting $\alpha_k=0$ and $w=1$, it is straightforward to show that~\eqref{eq:41} admits a unique NE. In what follows, we will demonstrate how the simplicity of $\mc X_i$ can facilitate algebraic manipulations of the KKT conditions of the best-response optimization problem~\eqref{maxi:op1} that will allow us to derive a semi-analytical method to compute it. 

We start by analyzing the stationarity conditions. Namely, for every player $i\in\mc I$, the Lagrangian of its best-response optimization problem simplifies to $\mc L_i\bigl(\mathbf{x}_i,\mathbf{x}_{-i},\lambda_i,\nu_i\bigr)=-\mathbf{u}_i(\mathbf{x}_i,\mathbf{x}_{-i})-\sum_{k\in\Z_K} \lambda_{i,k}x_{i,k}+\nu_i(\mathbf{1}^T\mathbf{x}_i-\mathbf{R}_i)$. Hence, for every $i\in\mc I$ and every $k\in\Z_K$, the Nash equilibrium $\overline{\mathbf{x}}\in\mc X$ has to satisfy for some feasible $\overline{\nu}_i\in\R$ and $\overline{\lambda}_{i}\in\R_+^K$
\begin{equation}\label{eq:42}
    \overline{\nu}_i-\overline{\lambda}_{i,k}+\beta_k-\frac{W_k(\sum_{j\neq i}\overline{x}_{j,k}+\varepsilon_k)}{(\sum_{j\in\mc I}\overline{x}_{j,k}+\varepsilon_k)^2}=0\,,
\end{equation}
which, after rearranging, gives for every $i\in\mc I$ and $k\in\Z_K$:
\begin{equation}\label{eq:43}
    \sum_{j\neq i}\overline{x}_{j,k}=\frac{1}{W_k}\bigl(\overline{\nu}_i-\overline{\lambda}_{i,k}+\beta_k\bigr)\biggl(\sum_{j\in\mc I}\overline{x}_{j,k}+\varepsilon_k\biggr)^2-\varepsilon_k\,.
\end{equation}
If we now sum up~\eqref{eq:43} for all $i\in\mc I$, we get for every $k\in\Z_K$
\begin{equation}\label{eq:44}
\begin{split}
    (N-1)&\sum_{j\in\mc I}\overline{x}_{j,k}= \\ &\sum_{i\in\mc I}\frac{1}{W_k}\bigl(\overline{\nu}_i-\overline{\lambda}_{i,k}+\beta_k\bigr)\biggl(\sum_{j\in\mc I}\overline{x}_{j,k}+\varepsilon_k\biggr)^2-N\varepsilon_k\,
    \end{split}
\end{equation}
which reads as 
\begin{equation}\label{eq:45}
    \Delta_k\overline{t}_k^2-W_k(N-1)\overline{t}_k-W_k\varepsilon_k=0
\end{equation}
for $\overline{t}_k\defineas\Phi_{k,\overline{\mathbf{x}}}+\varepsilon_k$, $\Delta_k\defineas N\beta_k+\mathbf{t}_{\nu}-\overline{\lambda}_{k}$, $\mathbf{t}_{\nu}\defineas\sum_{i\in\mc I}\overline{\nu}_i$, and $\overline{\lambda}_k\defineas\sum_{i\in\mc I}\overline{\lambda}_{i,k}$. Due to the specific structure of $\mc X_i$, it is clear that $\overline{t}_k>0$ for all $k\in\Z_K$. Hence,~\eqref{eq:45} will have a solution $\overline{t}_k>0$ if and only if $\Delta_k>0$, and is given by
\begin{equation}\label{eq:46}
    \overline{t}_k\defineas\frac{1}{2\Delta_k}\biggl(\overline{\mathbf{W}}_k+\sqrt{\overline{\mathbf{W}}_k^2+4W_k\Delta_k\varepsilon_k}\biggr)\,,
\end{equation}
for $\overline{\mathbf{W}}_k\defineas W_k(N-1)$. Let us now define  $\Sigma_\varepsilon\defineas\sum_{k\in\Z_K}\varepsilon_k$ and $\alpha_k\defineas N\beta_k-\overline{\lambda}_{k}$. Then, after summing~\eqref{eq:46} for all $k\in\Z_K$, we can define a function $f:\R\rightarrow\R$ by
\begin{multline}\label{eq:47}
    f(\mathbf{t}_{\nu})=\sum_{k\in\Z_K}\frac{\overline{\mathbf{W}}_k+\sqrt{\overline{\mathbf{W}}_k^2+4W_k\varepsilon_k(\alpha_k+\mathbf{t}_{\nu})}}{2(\alpha_k+\mathbf{t}_{\nu})}-\\-\sum_{i\in\mc I}\mathbf{R}_i-\Sigma_{\varepsilon}\,,
\end{multline}
\normalsize
making $\mathbf{t}_{\nu}$ a zero of $f$, i.e., $f(\mathbf{t}_{\nu})=0$. The problem now essentially boils down to finding the zeros of $f$, which will represent the backbone for characterizing the NE. Hence, we first prove the following lemma. 
\begin{lemma}\label{lema:6}
    Let the Blotto game $\mc G(K,\mc I \:|\: \mc E)$ be defined by~\eqref{eq:41}. If for every $i\in\mc I$, $\{\lambda_{i,k}\}_{k\in\Z_K}$ denotes an a priori chosen set of parameters such that $\lambda_{i,k}\geq 0$, then there exists at most one $\overline{\mathbf{x}}\in\mc X$ satisfying the stationarity condition~\eqref{eq:42}.
\end{lemma}
\medskip
\begin{proof}
    The domain of function $f(\mathbf{t}_{\nu})$ given by~\eqref{eq:47} is  $\mc D=\bigl[\max_{k}-\alpha_k-\frac{W_k(N-1)^2}{4\varepsilon_k},\infty\bigr)\setminus\{\alpha_k\}_{k\in\Z_K}$. Observe that, $f(\mathbf{t}_{\nu})$ has a vertical asymptote at $-\alpha_k$ for every $k\in\Z_K$ since $\lim_{\mathbf{t}_\nu\rightarrow-\alpha_k^+}f(\mathbf{t}_\nu)=\infty$ and $\lim_{\mathbf{t}_\nu\rightarrow-\alpha_k^-}f(\mathbf{t}_\nu)=-\infty$. Moreover, $\lim_{\mathbf{t}_\nu\rightarrow\infty}f(\mathbf{t}_\nu)=-\sum_{i\in\mc I}\mathbf{R}_i-\mathbf{\Sigma}_{\varepsilon}$. If $-\tilde{\alpha}_0$, $-\tilde{\alpha}_1\hdots-\tilde{\alpha}_{K-1}$ represent a non-decreasing permutation of $-\alpha_j$ for all $j\in\Z_K$, then Bolzano's theorem~\cite[Th.1]{Cauchy} ensures that there exists at least one zero of $f(\mathbf{t}_\nu)$ in each of the intervals $(-\tilde{\alpha}_{K-1},\infty)$ and $(-\tilde{\alpha}_k,-\tilde{\alpha}_{k+1})$ for $0\leq k\leq K-2$. An illustration plot of $f(\mathbf{t}_\nu)$ for $j\in\Z_3$ is shown in Figure~\ref{fig:1}.
    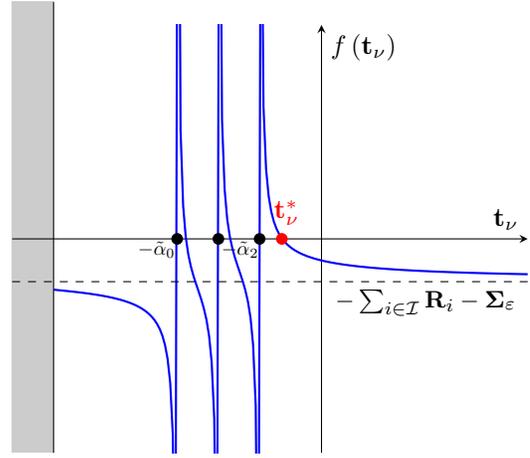
\begin{figure}[t]
    \centering
    \begin{tikzpicture}
        \begin{axis}[
            xlabel={$\mathbf{t}_{\nu}$},
            ylabel={$f\left(\mathbf{t}_{\nu}\right)$},
            domain=-15:10,
            samples=200,
            axis lines=middle,
            grid,
            ymin=-15, ymax=15, 
            xtick={\empty},
            ytick={\empty},
        ]
            \addplot[blue,thick] {(1+(1+0.1*(3+x))^0.5)/(3+x)+(1+(1+0.1*(5+x))^0.5)/(5+x)+(1+(1+0.1*(7+x))^0.5)/(7+x)-3};
            \addplot[black,dashed]{-3};
            \addplot[red,mark=*] coordinates {(-1.922,0)};
            \addplot[black,mark=*] coordinates {(-3,0)};
            \addplot[black,mark=*] coordinates {(-5,0)};
            \addplot[black,mark=*] coordinates {(-7,0)};   
            
        \end{axis}
        \node[scale=0.75](n1) at (1.92,2.7) {$-\tilde{\alpha}_0$};
        \node[scale=0.75](n2) at (3.03,2.7) {$-\tilde{\alpha}_2$};

        \node[red](n4) at (3.65,3.2) {$\mathbf{t}_{\nu}^*$};
        \node[black, ](n5) at (5.5,2) {\small$-\sum_{i\in\mc I}\mathbf{R}_i-\mathbf{\Sigma}_{\varepsilon}$};

        \fill[black, fill opacity=0.2] (0.55,0) rectangle (0,6);
        \draw[](0.55,0) -- (0.55,6);

    \end{tikzpicture}
    \caption{\unboldmath Illustrative example plot of the function~\eqref{eq:47} for $k\in\Z_3.$}
    \label{fig:1}
\end{figure}
Since $\Delta_k>0$ for all $k\in\Z_K$, it is clear that all the zeros of $f(\mathbf{t}_{\nu})$ must satisfy $\mathbf{t}_{\nu}^*\geq\max_{k}-\alpha_k=-\tilde{\alpha}_{K-1}$. Hence, all feasible zeros must be in the interval $(-\tilde{\alpha}_{K-1},\infty)$. However, by looking at the first derivative $f'(\mathbf{t}_{\nu})$ given by
\begin{equation}
\begin{split}
    f'&(\mathbf{t}_{\nu})=  \\ & \sum_{k\in\Z_K}\frac{1}{2(\alpha_k+\mathbf{t}_{\nu})^2}\biggl(\frac{-\overline{\mathbf{W}}_k^2-2W_k\varepsilon_k(\alpha_k+\mathbf{t}_{\nu})}{\sqrt{\overline{\mathbf{W}}_k^2+4W_k\varepsilon_k(\alpha_k+\mathbf{t}_{\nu})}}-\overline{\mathbf{W}}_k\biggr)\,,
    \label{eq:48}
    \end{split}
\end{equation}
and observing that $\alpha_k+\mathbf{t}_{\nu}\geq -\frac{W_k}{4\varepsilon_k}(N-1)^2$ due to the lower bound of $\mc D$, we have that $-\overline{\mathbf{W}}_k^2-2W_k\varepsilon_k(\alpha_k+\mathbf{t}_{\nu})\leq -\overline{\mathbf{W}}_k^2+\frac{1}{2}\overline{\mathbf{W}}_k^2<0$, resulting in $f'(\mathbf{t}_{\nu})<0$. This means that $f$ is strictly decreasing on $(-\tilde{\alpha}_{K-1},\infty)$ which combined with Bolzano's theorem~\cite[Th.1]{Cauchy}, ensures there is exactly one zero $\mathbf{t}^*_{\nu}$ on $(-\tilde{\alpha}_{K-1},\infty)$. By plugging $\mathbf{t}^*_{\nu}$ into $\Delta_k$ and using it in~\eqref{eq:46} to obtain $\overline{t}_k^*$, we can extract $\overline{x}_{i,k}$ from~\eqref{eq:43} as
\begin{equation}\label{eq:50}
    \overline{x}_{i,k}=\overline{t}_k^*-\frac{1}{W_k}(\overline{\nu}_i-\lambda_{i,k}+\beta_k)\left.\overline{t}_k^*\right.^2\,.
\end{equation}
If we now sum this over all the stages $k\in\Z_k$, we get
\begin{equation}\label{eq:51}
    \overline{\nu}_i=\frac{\sum_{k\in\Z_k}\overline{t}_k^*-\mathbf{R}_i-\sum_{k\in\Z_k}(-\lambda_{i,k}+\beta_k)\frac{\left.\overline{t}_k^*\right.^2}{W_k}}{\sum_{k\in\Z_k}\frac{\left.\overline{t}_k^*\right.^2}{W_k}}\,,
\end{equation}
which directly determines the unique values of all $\overline{x}_{i,k}$ given by~\eqref{eq:50}. If all of them are non-negative, then they correspond to the unique $\overline{\mathbf{x}}\in\mc X$ satisfying~\eqref{eq:42}. Otherwise, for this choice of $\{\lambda_{i,k}\}_{k\in\Z_k}$ such $\overline{\mathbf{x}}\in\mc X$ does not exist.   
\end{proof}
\medskip
Note that a tuple $\bigl(\overline{\mathbf{x}}, \{\lambda_{i,k}\}_{i\in\mc I,k\in\Z_K},\{\overline{\nu}_i\}_{i\in\mc I}\bigr)$, given by~\eqref{eq:50} and~\eqref{eq:51}, will be a solution of the best-response optimization problem~\eqref{maxi:op1} if the complementary slackness condition also holds. Namely, if $\lambda_{i,k}>0$ for some $i\in\mc I$ and $k\in\Z_K$, then it also has to hold that $\overline{x}_{i,k}=0$. Conversely, if $\overline{x}_{i,k}>0$, then $\lambda_{i,k}=0$ has to hold. Since $\mathbf{t}^*_{\nu}$ has to be found numerically and the choice of $\{\lambda_{i,k}\}_{i\in\mc I,k\in\Z_K}$ directly determines the interval $(-\overline{\alpha}_{K-1},\infty)$ in which $\mathbf{t}^*_{\nu}$ is located, finding the NE is in general hard as it requires exploration of the unbounded space of dual variables $\lambda_{i,k}$. However, if the unique Nash equilibrium is an interior point of $\mc X$, i.e., $\overline{x}_{i,k}>0$ for all $i\in\mc I$ and $k\in\Z_K$, then~\eqref{eq:50} and~\eqref{eq:51} provide its analytical form when $\lambda_{i,k}$ is set to zero for all $i\in\mc I$ and $k\in\Z_K$. We can now leverage this to find the NE in a semi-analytical manner regardless of its position by examining all possible configurations in which certain (if any) battlefield participations are fixed to zero. 

Formally speaking, if $\mc P\defineas 2^{\mc I\times \Z_K}$ represents the power set of $\mc I\times \Z_K$, then any $\mc O\in\mc P$ defines one configuration in which some players (if any) do not participate in certain battlefields,
i.e., $\overline{x}_{i,k}=0$ for every $(i,k)\in\mc O$. It is important to note that some $\mc O\in\mc P$ cannot satisfiy the constraint $\mathbf{1}_K^T\mathbf{x}_i=\mathbf{R}_i$. Specifically, for each $i\in\mc I$, not all pairs $(i,k)$, with $k\in\Z_K$, can be simultaneously included in a feasible $\mc O$. Hence, if
\begin{align}
    \overline{\mc O}_k&\defineas\{i\in\mc I\:\mid\:(i,k)\notin\mc O\}\quad\quad\text{for all }k\in\Z_K, \label{eq:o1}\\
    \underline{\mc O}_i&\defineas\{k\in\Z_K\:\mid\:(i,k)\in\mc O\}\quad\text{for all }i\in\mc I, 
\end{align}
represent the $\mc O$-induced sets for any $\mc O\in\mc P$, then the set of all feasible configurations is given by 
\begin{equation}\label{eq:f}
    \mc F\defineas\{\mc O\in\mc P\:\mid\:|\underline{\mc O}_i|<K\text{ for all }i\in\mc I\}.
\end{equation}
If we now iterate over all $\mc O\in\mc F$ and apply the procedure for finding the interior point Nash equilibrium based on~\eqref{eq:50} and~\eqref{eq:51}, we are certain to find the unique NE of $\mc G(K,\mc I \:|\: \mc E)$. This is summarized in the following theorem.
\begin{theorem}\label{th:3}
    Let the Blotto game $\mc G(K,\mc I \:|\: \mc E)$ be defined by~\eqref{eq:41}, $\mc P\defineas 2^{\mc I\times\Z_K}$, and the set $\mc F$ be given by~\eqref{eq:f}. Then, the unique Nash equilibrium $\overline{\mathbf{x}}\in\mc X$ of $\mc G(K,\mc I \:|\: \mc E)$ is given by a unique configuration $\mc O\in\mc F$ inducing the sets $\overline{\mc O}_k$ given by~\eqref{eq:o1}, with $\mathbf{n}_k=|\overline{\mc O}_k|$ for every $k\in\Z_K$, such that the following description holds: 
    \begin{align}
        \overline{x}_{i,k}&=\begin{cases}
        0, & \text{if } (i,k)\in\mc O \\ 
        \overline{t}_k^*-\frac{1}{W_k}(\overline{\nu}_i+\beta_k)\left.\overline{t}_k^*\right.^2, & \text{otherwise} 
        \end{cases} \label{eq:53:1}\\        \overline{\nu}_i&=\frac{\sum_{k\in\Z_k}\overline{t}_k^*-\mathbf{R}_i-\sum_{k\in\Z_k}\beta_k\frac{\left.\overline{t}_k^*\right.^2}{W_k}}{\sum_{k\in\Z_k}\frac{\left.\overline{t}_k^*\right.^2}{W_k}} \label{eq:53:2}\\  \overline{t}_k^*&=\frac{\widetilde{\mathbf{W}}_k+\sqrt{\widetilde{\mathbf{W}}_k^2+4W_k\varepsilon_k(\mathbf{t}_{\nu,k}^*+\mathbf{n}_k\beta_k)}}{2(\mathbf{t}_{\nu,k}^*+\mathbf{n}_k\beta_k)} \label{eq:53:4}
    \end{align}
    with $\widetilde{\mathbf{W}}_k=W_k(\mathbf{n}_k-1)$ and $\mathbf{t}_{\nu,k}^*$ being the unique solution of $\tilde{f}(\mathbf{t}_{\nu,k})=0$ given by
    \begin{multline}\label{eq:52}
        \tilde{f}(\mathbf{t}_{\nu,k})=\sum_{k\in\Z_K}\frac{\widetilde{\mathbf{W}}_k+\sqrt{\widetilde{\mathbf{W}}_k^2+4W_k\varepsilon_k(\mathbf{t}_{\nu,k}+\mathbf{n}_k\beta_k)}}{2(\mathbf{t}_{\nu,k}+\mathbf{n}_k\beta_k)}-\\-\sum_{i\in\mc I}\mathbf{R}_i-\Sigma_{\varepsilon}\,,
    \end{multline}
    such that $\mathbf{t}_{\nu,k}^*\in\bigl[\max_{k}-\mathbf{n}_k\beta_k,\infty\bigr)$.   
\end{theorem}
\medskip
\begin{proof}
    Since there is a unique NE of $\mc G(K,\mc I \:|\:  \mc E)$, there exists exactly one configuration $\mc O\in\mc P$ describing it. For every $(i,k)\in\mc I\times\Z_K\setminus\mc O$, the complementary slackness yields $\lambda_{i,k}=0$, which results in $\sum_{j\neq i}\overline{x}_{j,k}=\frac{1}{W_k}\bigl(\overline{\nu}_i+\beta_k\bigr)(\Phi_{k,\overline{\mathbf{x}}}+\varepsilon_k)^2-\varepsilon_k$. If for every $k\in\Z_K$, we sum this up over $i\in\overline{\mc O}_k$, we first get $$\sum_{i\in\overline{\mc O}_k}\sum_{j\neq i}\overline{x}_{j,k}=\mathbf{n}_k\Phi_{k,\overline{\mathbf{x}}}-\sum_{i\in\overline{\mc O}_k}\overline{x}_{i,k}=(\mathbf{n}_k-1)\Phi_{k,\overline{\mathbf{x}}},$$ since $\overline{x}_{i,k}=0$ for $(i,k)\in\mc O$. Therefore, we have
    \begin{equation}
        \label{eq:53}
        (\mathbf{n}_k-1)\Phi_{k,\overline{\mathbf{x}}}=\sum_{i\in\overline{\mc O}_k}\frac{1}{W_k}\bigl(\overline{\nu}_i+\beta_k\bigr)\biggl(\Phi_{k,\overline{\mathbf{x}}}+\varepsilon_k\biggr)^2-\mathbf{n}_k\varepsilon_k\,
    \end{equation}
    \normalsize
    which after rearranging gives the equivalent form of~\eqref{eq:45} if we substitute $\overline{\lambda}_k\rightarrow 0$, $N\rightarrow \mathbf{n}_k$, and $\mathbf{t}_{\nu}\rightarrow \mathbf{t}_{\nu,k}=\sum_{i\in\overline{\mc O}_k}\overline{\nu}_i$. Therefore, we can directly invoke the analysis done in Lemma~\ref{lema:6} to conclude the proof.     
\end{proof}
\medskip
From the implementation standpoint, the computational complexity of Theorem~\ref{th:3} stems from having to compute the solution of $\tilde{f}(\mathbf{t}_{\nu,k})=0$ for every $k\in\Z_K$. Since we already know the interval in which the solution should be located and $\tilde{f}$ is monotone, we can use fast, off-the-shelf solvers to obtain $\tilde{\mathbf{t}}^*_{\nu,k}$. Nevertheless, the same set of steps needs to be repeated until a feasible configuration $\mc O\in\mc F$ that ensures $\overline{x}_{i,k}>0$ for all $(i,k)\in\mc I\times\Z_K \setminus\mc O$ has been found. In practice, for various configurations of game parameters, it is to be expected that the Nash equilibrium resides in the interior of the feasible sets. However, for setups where this is not the case, the limited scalability of the method based on Theorem~\ref{th:3} can result in a longer computation time compared to Algorithm~\ref{al:1}. To compare the temporal complexity of the two approaches, we recall that the main technical challenges in Algorithm~\ref{al:1} lie in the need to solve the optimization problems for the optimality check and the projection step. Since this typically involves using off-the-shelf optimization libraries, the execution time might vary based on the software architecture used.
Let us assume that the variable assignment steps related to updating players' decision variables take one time instance to execute, while the projection, optimality check, and solving $\tilde{f}(\mathbf{t}_{\nu,k})=0$ take $\tau_{\text{proj}}$, $\tau_{\text{opt}}$ and  $\tau_{\text{zero}}$ instances, respectively. Then, the following corollary roughly compares the temporal complexity of the two algorithms. 
\begin{corollary}
     Consider a Blotto game $\mc G(K,\mc I \:|\: \mc E)$ as defined in ~\eqref{eq:41}  with an $L$-Lipshitz and $\mu$-strongly monotone pseudo-gradient. Let Algorithm~\ref{al:1} be initialized with $\overline{\gamma}>0$, $\eta\in(0,1)$, and a sufficiently large inner-loop timeout parameter $t_{\text{out}}$. Then, the temporal complexity of Algorithm~\ref{al:1} is
    \begin{equation}\label{eq:54}
        \mathbf{T}_{\text{alg},1}=\biggl(1+\log_{\eta}\frac{2\mu}{L^2\overline{\gamma}}\biggr)\bigl(\tau_{\text{opt}}+(K+\tau_{\text{proj}})t_{\text{out}}\bigr)\,,
    \end{equation}
    and the temporal complexity of the semi-analytical method is 
    \begin{equation}\label{eq:55}
        \mathbf{T}_{\text{alg},2}=(2^K-1)^N\bigl(K+1+K(\tau_{\text{zero}}+2)\bigr)\,.
    \end{equation}     
\end{corollary}
\medskip
\begin{proof}
    We start by noting that at each iteration of Algorithm~\ref{al:1}'s outer loop, we have to perform $K$ assignments to perform $\mathbf{x}_i^{(t)}+\gamma \nabla_{\mathbf{x}_i}\mathbf{u}_i(\mathbf{x}_i^{(t)},\mathbf{x}_{-i}^{(t)})$, before one projection step is performed. Although this can be performed in parallel for all $i\in\mc I$, in the worst case scenario, this set of steps is repeated $t_{\text{out}}$ times, after which the optimality test is performed and one iteration of the outer loop can be concluded. Since the outer loop has to be repeated at most until $\eta^l\overline{\gamma}<\frac{2L}{\mu^2}$, i.e., until $l>\log_{\eta}\frac{2L}{\mu^2\overline{\gamma}}$ yielding $1+\log_{\eta}\frac{2L}{\mu^2\overline{\gamma}}$ iterations, the total temporal complexity is given by~\eqref{eq:54}. Similarly, for every feasible $\mc O\in\mc F$, we have to solve $K$ times $f(\mathbf{t}_{\nu,k})=0$ to compute the corresponding $\overline{t}_k^*$ and $\Phi_{k,\overline{\mathbf{x}}}^*$. Then, all players can in parallel compute the personal $\overline{\nu}_i$ and $K$ values $\overline{x}_{i,k}$. To determine the number of $\mc O\in\mc F$, we recall that for a particular $i\in\mc I$, every pair $(i,k)$ for $k\in\Z_K$ can, but does not necessarily have to, be included in $\mc O$, yielding $2^K$ combinations. However, a combination where all $(i,k)$ are included is not feasible as it directly implies $\mathbf{1}_K^T\mathbf{x}_i=0$. Therefore, for each $i\in\mc I$ we have $2^K-1$ possible combinations, resulting in $(2^K-1)^N$ feasible $\mc O\in\mc F$. Since the temporal complexity of each feasible $\mc O\in\mc F$ is given by $K(\tau_{\text{zero}}+2)+K+1$, the total temporal complexity of the second procedure is given by~\eqref{eq:55}.            
\end{proof}

Note that the temporal complexity given by $\mathbf{T}_{\text{alg},2}$ in~\eqref{eq:55} is a pessimistic overestimate. In practice, for a specific configuration $\mc O\in\mc F$ with the corresponding induced sets $\overline{\mc O}_k$, the zero-finding procedure only needs to be performed as many times as there are distinct values of $\mathbf{n}_k$. For instance, if the unique Nash equilibrium lies in the interior of the feasible set, then for all $k\in\Z_K$, we have $\mathbf{n}_k=|\mc I|$. In this case, the function $\tilde{f}(\mathbf{t}_{\nu,k})$ remains the same for all $k$, meaning that the zero-finding procedure needs to be executed only once. Also, the multiplicative term $(2^K-1)^N$ reflects the worst-case scenario, as the search can be terminated as soon as a Nash equilibrium is found. 
\begin {figure}
\centering
\begin{adjustbox}{max height=0.55\textwidth, max width=0.48\textwidth}
\begin{tikzpicture}[scale=1.0, font=\Large]

    \node[draw=mycolor1!10, rectangle, rounded corners=2.5mm, minimum width=4.5cm, minimum height=3.5cm, anchor=center, fill=mycolor1!10](2b2) at (-4,0){};
    \node[businessman, shirt=mycolor1, scale=3.5](2b1) at (-4.0, 0.25){};
    \node[ text=black, shape=rectangle,scale=1.75](ch1) at (-4.0, -1.25){\textbf{Company $1$}};

    \node[draw=mycolor2!10, rectangle, rounded corners=2.5mm, minimum width=4.5cm, minimum height=3.5cm, anchor=center, fill=mycolor2!10](2b2) at (3,0){};
    \node[businessman, shirt=mycolor2, scale=3.5](2b2) at (3,0.25){};
    \node[ text=black, shape=rectangle,scale=1.75](ch1) at (3, -1.25){\textbf{Company $2$}};

    \node[draw=purple!10, rectangle, rounded corners=2.5mm, minimum width=4.5cm, minimum height=3.5cm, anchor=center, fill=purple!10](2b2) at (10,0){};
    \node[businessman, shirt=purple, scale=3.5](2b3) at (10,0.25){};
    \node[ text=black, shape=rectangle,scale=1.75](ch1) at (10.0, -1.25){\textbf{Company $3$}};

    \node[scale=1.25] (pic1) at (-12, -9) {\includegraphics[width=.35\textwidth]{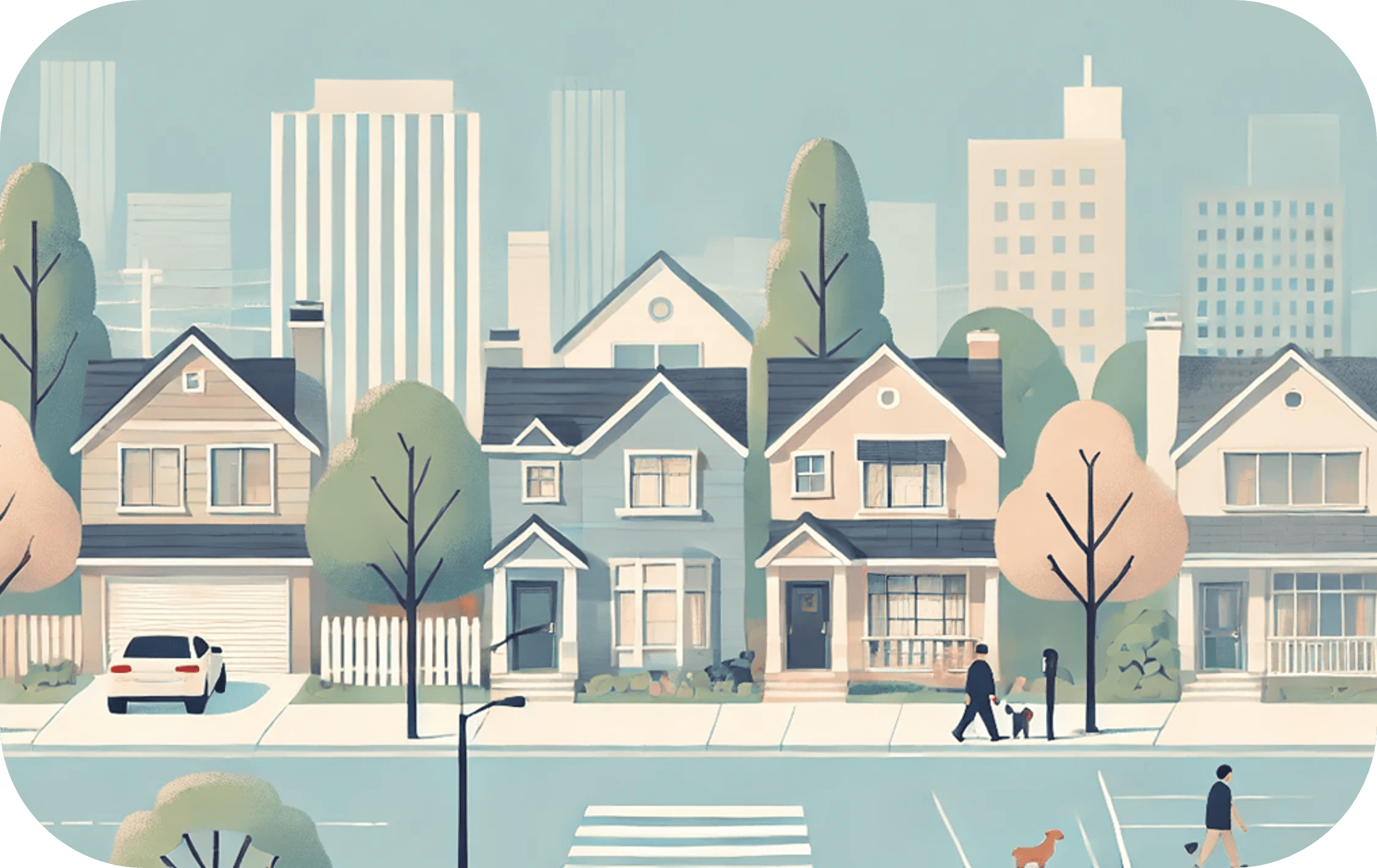}};    
    \node[ text=black, shape=rectangle,scale=1.75](ch1) at (-12.0, -12){\textbf{Region $J_1$}};

    \node[scale=1.25] (pic2) at (-2, -9) {\includegraphics[width=.35\textwidth]{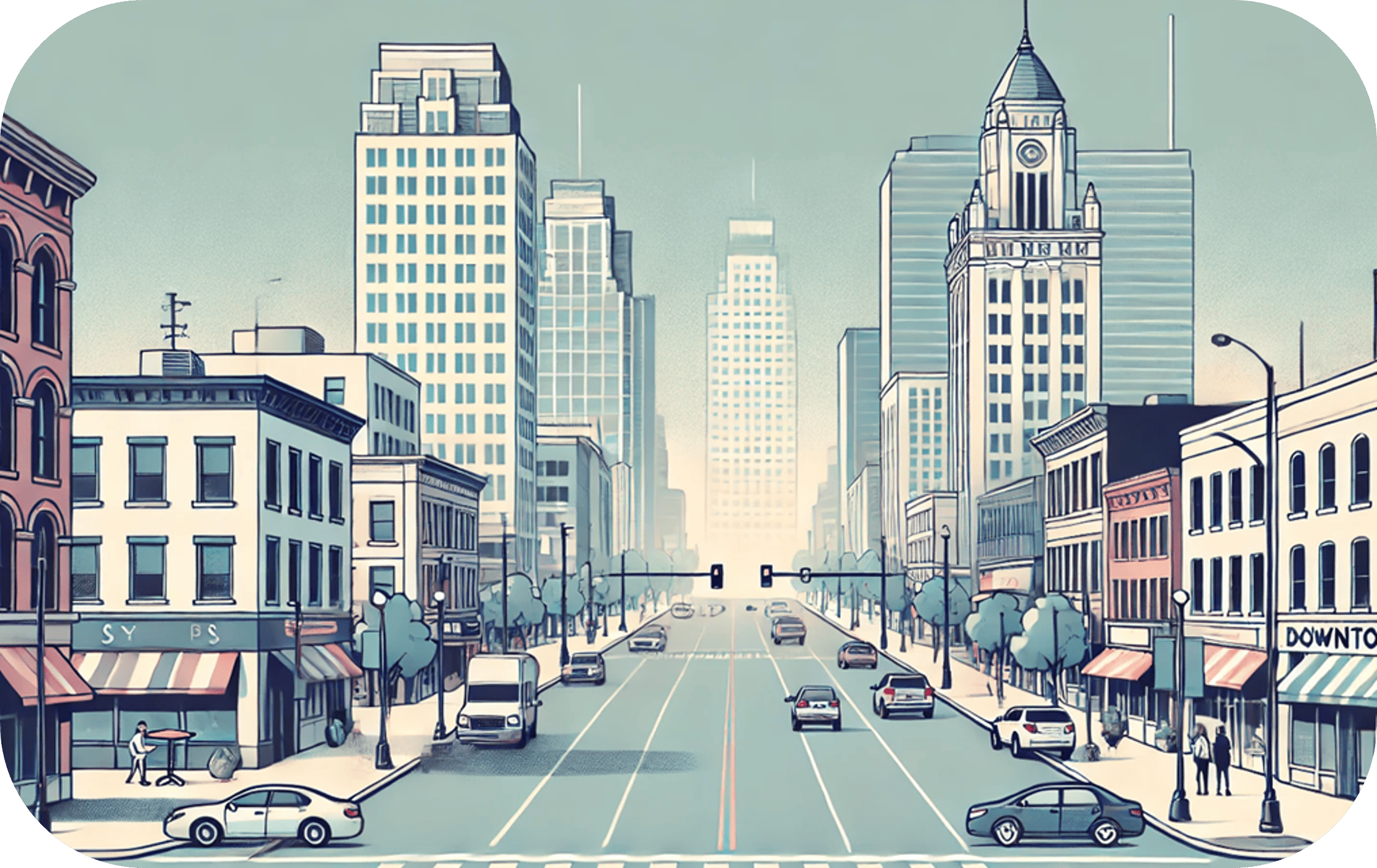}};
    \node[ text=black, shape=rectangle,scale=1.75](ch1) at (-2.0, -12){\textbf{Region $J_2$}};
    
    \node[scale=1.25] (pic3) at (8, -9) {\includegraphics[width=.35\textwidth]{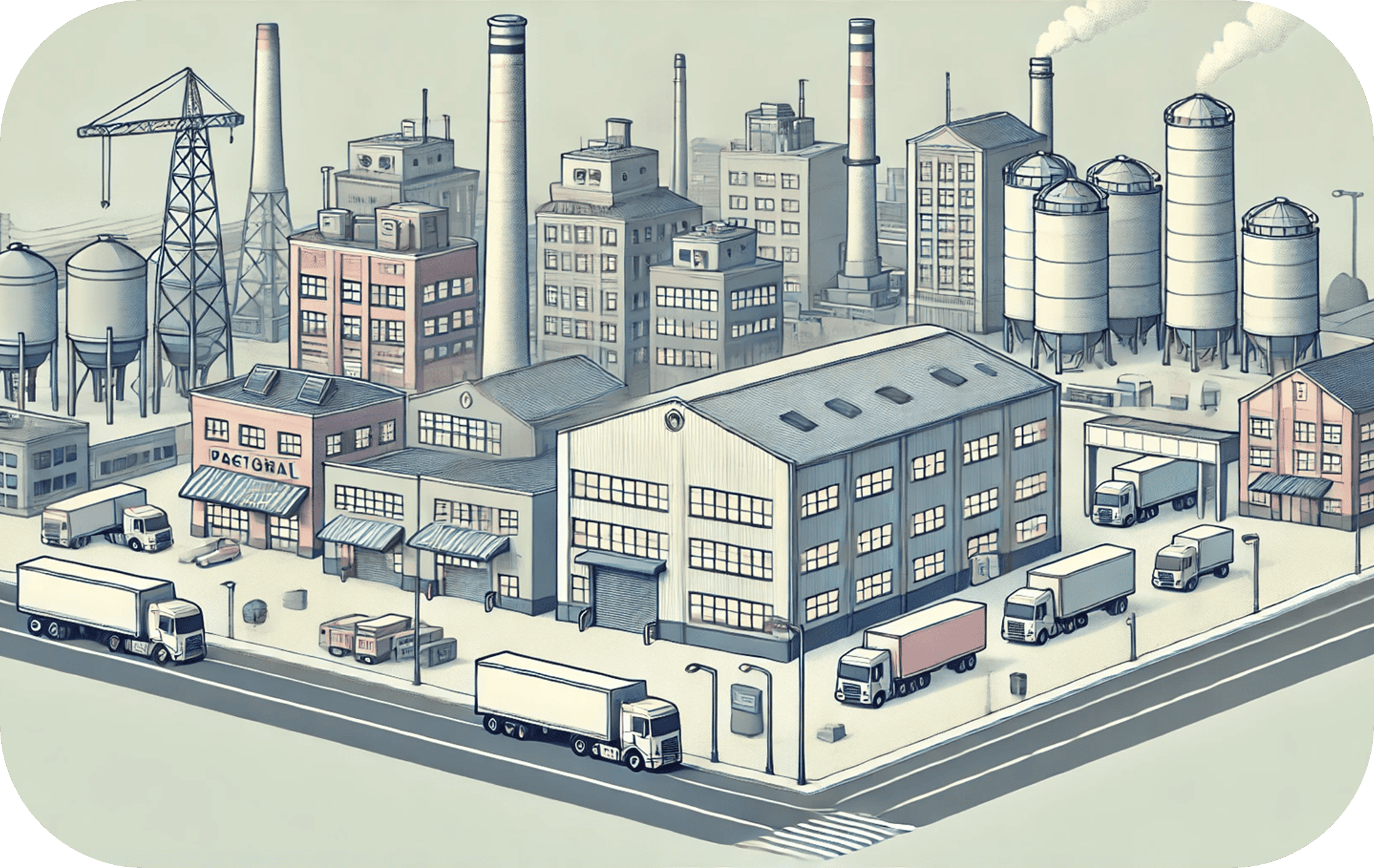}};
    \node[ text=black, shape=rectangle,scale=1.75](ch1) at (8.0, -12){\textbf{Region $J_3$}};
    
    \node[scale=1.25] (pic4) at (18, -9) {\includegraphics[width=.35\textwidth]{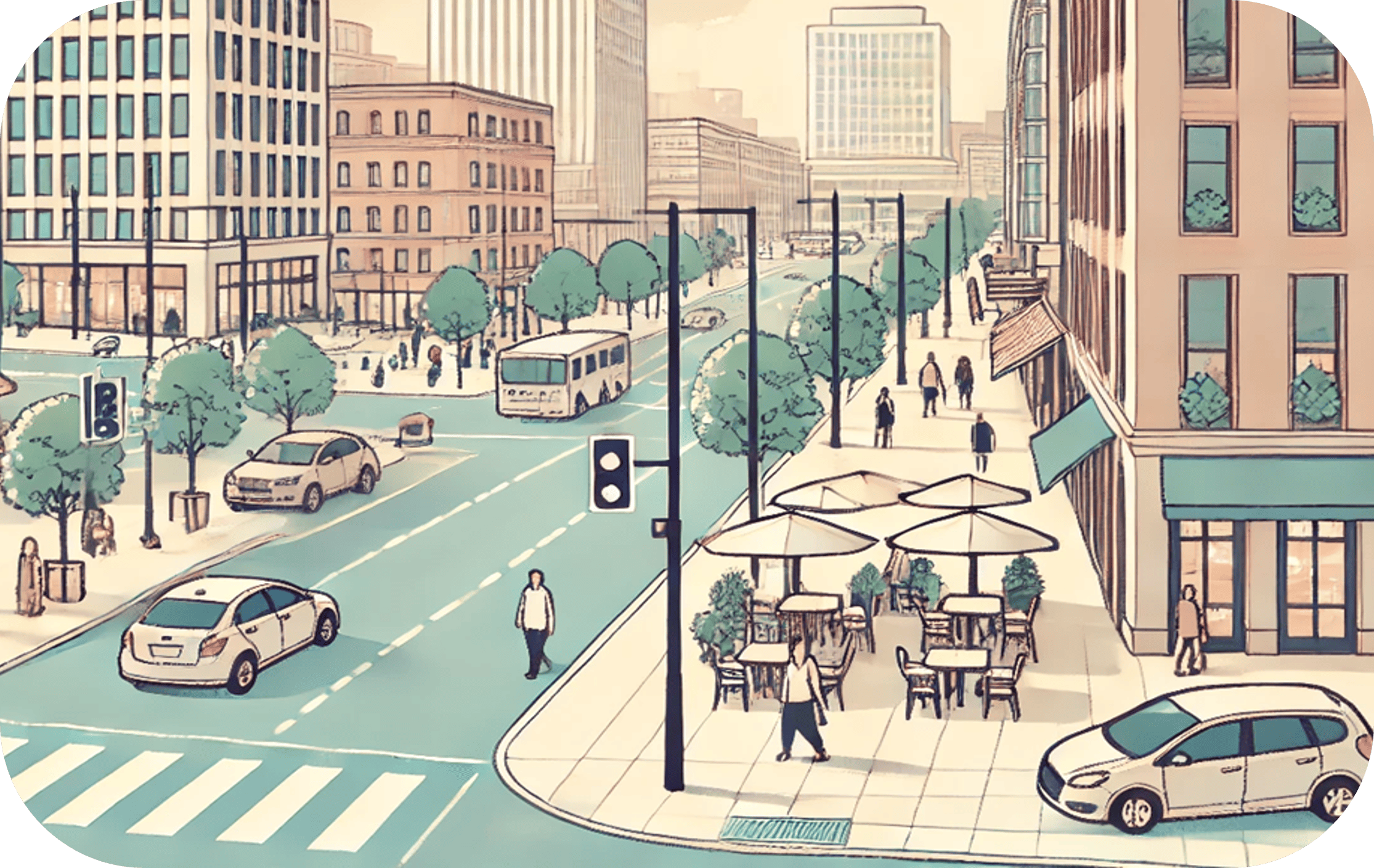}};
    \node[ text=black, shape=rectangle,scale=1.75](ch1) at (18.0, -12){\textbf{Region $J_4$}};

    \node[scale=1.6] (pic5) at (-10.0,-21.7) {\usepgfplotslibrary{groupplots}

\begin{tikzpicture}

\definecolor{color0}{rgb}{1,0.498039215686275,0.0549019607843137}
\definecolor{color1}{rgb}{0.580392156862745,0.403921568627451,0.741176470588235}

\begin{groupplot}[group style={group size=1 by 3}]
\nextgroupplot[
legend cell align={right},
legend style={fill opacity=0.8, draw opacity=1, text opacity=1, draw=white!90!black,at={(1,1)}, font=\small},
tick align=outside,
tick pos=left,
x grid style={white!90!black},
xmajorgrids,
xmin=-0.1, xmax=9.1,
xtick style={color=black},
y grid style={white!90!black},
ylabel=\textcolor{black}{$W_k^{1,\text{lo}}$},
ymin=-2750, ymax=80000,
ytick style={color=black},
height=2.7cm,
width=8cm,
ytick={0, 40000, 80000},
]
\addplot [semithick, black, const plot mark left, dashed,  mark size=3, mark options={solid}]
table {%
0 2500
1 10000
2 30000
3 65000
4 45000
5 30000
6 20000
7 12500
8 5000
9 5000
};

\nextgroupplot[
legend cell align={right},
legend style={fill opacity=0.8, draw opacity=1, text opacity=1, draw=white!90!black,at={(1,1)}, font=\small},
tick align=outside,
tick pos=left,
x grid style={white!90!black},
xmajorgrids,
xmin=-0.1, xmax=9.1,
xtick style={color=black},
y grid style={white!90!black},
ylabel=\textcolor{black}{$\alpha_k$},
ylabel style={xshift=-4pt},
ymin=0.0, ymax=1.6,
ytick style={color=black},
height=2.7cm,
width=8cm,
ytick={0, 1, 2},
]
\addplot [semithick, black, const plot mark left, dashed,  mark size=3, mark options={solid}]
table {%
0 1
1 1
2 0.1
3 0.1
4 0.1
5 0.5
6 1.5
7 1.5
8 1.5
9 1.5
};

\nextgroupplot[
legend cell align={right},
legend style={fill opacity=0.8, draw opacity=1, text opacity=1, draw=white!90!black,at={(1,1)}, font=\small},
tick align=outside,
tick pos=left,
x grid style={white!90!black},
xlabel={Time interval $k$},
xmajorgrids,
xmin=-0.1, xmax=9.1,
xtick style={color=black},
y grid style={white!90!black},
ylabel=\textcolor{black}{$\varepsilon_k$},
ylabel style={xshift=-4pt},
ymin=8, ymax=52,
ytick style={color=black},
height=2.7cm,
width=8cm,
ytick={10, 30, 50},
]
\addplot [semithick, black, const plot mark left, dashed,  mark size=3, mark options={solid}]
table {%
0 10
1 20
2 30
3 50
4 50
5 40
6 20
7 10
8 10
9 10
};

\end{groupplot}

\end{tikzpicture}};
    \draw[shift={(-9.5, -20.5)},rounded corners=5mm] (-7,-7) rectangle (7,5);

    \node[scale=0.7] (pic6) at (10.5, -21.5){\input{figures/RHG.tikz}};
    \draw[shift={(10.5, -20.5)},rounded corners=5mm] (-12,-7) rectangle (12,5);

    \draw[-{Triangle[length=8pt, width=7.5pt]},line width=1.8pt, draw=mycolor1](-4, -1.75)--(pic1.north);
    \draw[-{Triangle[length=8pt, width=7.5pt]},line width=1.8pt, draw=mycolor1](-4, -1.75)--(pic2.north);
    \draw[-{Triangle[length=8pt, width=7.5pt]},line width=1.8pt, draw=mycolor1](-4, -1.75)--(pic3.north);    
    \draw[-{Triangle[length=8pt, width=7.5pt]},line width=1.8pt, draw=mycolor1](-4, -1.75)--(pic4.north); 

    \draw[-{Triangle[length=8pt, width=7.5pt]},line width=1.8pt, draw=mycolor2](3, -1.75)--(pic1.north);
    \draw[-{Triangle[length=8pt, width=7.5pt]},line width=1.8pt, draw=mycolor2](3, -1.75)--(pic2.north);
    \draw[-{Triangle[length=8pt, width=7.5pt]},line width=1.8pt, draw=mycolor2](3, -1.75)--(pic3.north);    
    \draw[-{Triangle[length=8pt, width=7.5pt]},line width=1.8pt, draw=mycolor2](3, -1.75)--(pic4.north);

    \draw[-{Triangle[length=8pt, width=7.5pt]},line width=1.8pt, draw=purple](10, -1.75)--(pic1.north);
    \draw[-{Triangle[length=8pt, width=7.5pt]},line width=1.8pt, draw=purple](10, -1.75)--(pic2.north);
    \draw[-{Triangle[length=8pt, width=7.5pt]},line width=1.8pt, draw=purple](10, -1.75)--(pic3.north);    
    \draw[-{Triangle[length=8pt, width=7.5pt]},line width=1.8pt, draw=purple](10, -1.75)--(pic4.north);

    \draw[line width=1.8pt, draw=black, dashed](-16, -11.05)--(-16.5, -15.75);
    \draw[line width=1.8pt, draw=black, dashed](-8,-11.05)--(-2.5, -15.75);

   \draw[-{Triangle[length=8pt, width=7.5pt]},line width=1.8pt, draw=black](-2.5, -21.5)--(-1.5, -21.5);

    \node[text=mycolor1, shape=rectangle,scale=2]() at (-7.4, -3){$x_{1,1}^{\text{up}}\phantom{x}$};
    \node[text=mycolor1, shape=rectangle,scale=2]() at (-4.1, -3){$x_{1,2}^{\text{up}}\phantom{x}$};

    \node[text=purple, shape=rectangle,scale=2]() at (10.2, -3){$\phantom{x}x_{3,3}^{\text{up}}$};
    \node[text=purple, shape=rectangle,scale=2]() at (13.2, -3){$\phantom{x}x_{3,4}^{\text{up}}$};

    \node[ text=black, shape=rectangle,scale=1.55](ch1) at (-9.5, -15){\textbf{Daily evolution of parameters in $J_1$}};
    \node[ text=black, shape=rectangle,scale=1.55](ch1) at (10.5, -15){\textbf{Receding horizon charging scheduling game $J_1$}};    
    
\end{tikzpicture}
\end{adjustbox}
    \caption{\unboldmath Illustration of the hierarchical operational management in a ride-hailing market with $|\mc I|=3$ companies. On the upper level, the companies aim to optimally distribute their fleets among regions $J=\{J_1,J_2,J_3,J_4\}$. Then, in each of the regions, they aim to design a charging scheduling scheme to take into account the inter-daily temporal variability of the region-specific parameters.} 
\label{fig:img_taccasestudy1}
\end{figure}

\section{Numerical Simulations}\label{sec:6}

We perform a numerical analysis of the proposed model within a simplified hierarchical framework that encapsulates the full operational management of transportation service providers in a future smart mobility market, as in~\cite{maljkovic2024blotto, maljkovic2024cdc}.

\subsection{Case study}\label{subsec:6a}

In particular, we look at a bi-level setup shown in Figure~\ref{fig:img_taccasestudy1}, where $|\mc I|=3$ companies operate electric vehicle fleets in a city area divided into $K_r=4$ regions. 

\textbf{Upper-level management:} Each company operator $i\in\mc I$ aims to optimally split its fleet of $\mathbf{R}_i\in\R_+$ vehicles among the $K_r$ regions, by taking into account the expected average profit from serving the demand $W_k^{\text{up}}\in\R_+$ and the average expected cost of charging a vehicle $\beta_k^{\text{up}}\in\R_+$ in each of the regions $j\in\{1,2,\ldots,K_r\}$. That is, each service provider aims to find a $\mathbf{x}_i^{\text{up}}\defineas\text{col}(x_{i,j}^{\text{up}})_{j\in\Z_{K_r}}$ such that $x_{i,j}^{\text{up}}\geq 0$, $\mathbf{1}^T\mathbf{x}_i^{\text{up}}=\mathbf{R}_i$, and the joint strategy $\mathbf{x}^{\text{up}}=\text{col}(\mathbf{x}_i^{\text{up}})_{i\in\mc I}$ represents a no-regret solution of the Blotto game $\mc G(K_r, \mc I \:|\:  \mc E)$ defined by~\eqref{eq:41}.

\textbf{Lower-level management:} Once the service provider operators split their vehicles into region-specific sub-fleets, their goal is to design a proactive charging scheduling scheme for each of the sub-fleets, so as to take into account the time-varying demand and charging prices during the day. Namely, for every region $j\in\{1,2,\ldots,K_r\}$, the operational management of the $x_{i,j}^{\text{up}}\in\R_+$ vehicles determined by the upper-level management is considered over a predefined horizon of $T=9$ time intervals. At every time interval $k\in\Z_{T}$, we assume that the $j$-th sub-fleet of company $i\in\mc I$ is described by a $b$-element aggregative state of charge (SoC) vector $\mathbf{y}_{i,k}^j\in\R_+^b$, such that $b\in\N$ represents the number of distinct battery level categories, and each element of this vector denotes how many of the $x_{i,j}^{\text{up}}$ vehicles belongs to a specific battery level category. If the input vector $\mathbf{u}_{i,k}^j\in\R_+^b$ represents how many vehicles of each battery level category are sent for charging at time interval $k\in\Z_T$, then the SoC evolves over the time horizon $T$ according to a simplified state-space model $\mathbf{y}_{i,k+1}^j=A_i^j\mathbf{y}_{i,k}^j+B_i^j\mathbf{u}_{i,k}^j$,  where the matrices $A_i^j,B_i^j\in\R^{b\times b}$ ensure that $\mathbf{1}^T\mathbf{y}_{i,k}^j=x_{i,j}^{\text{up}}$, and are explained in detail in~\cite{maljkovic2024cdc}. If we assume that vehicles in the lowest battery level category are either sent for recharging or remain parked, and that those in either state cannot serve demand during interval $k$, then the auxiliary market participation function of company $i\in\mc I$ in region $j$ at time interval $k$ in line with~\eqref{eq:auxpart} can be given by $\varphi_{i,k}(\mathbf{y}_{i,k}^{j},\mathbf{u}_{i,k}^{j})=\Lambda^T(\mathbf{y}_{i,k}^{j}-\mathbf{u}_{i,k}^{j})$, where $\Lambda$ is a selection matrix such that $\Lambda^T=[0,1,\ldots,1]$. Now, by concatenating this auxiliary participation function and the input vectors, we obtain a single decision variable
\begin{equation}
    \mathbf{x}_i^{j,\text{lo}}\defineas\text{col}(x_{i,k}^{j,\text{lo}})_{k\in\Z_T}\:\land\: x_{i,k}^{j,\text{lo}}=\bigl[\varphi_{i,k}(\mathbf{y}_{i,k}^j,\mathbf{u}_{i,k}^j),  (\mathbf{u}_{i,k}^j)^T\bigr]^T,
\end{equation}
suitable for defining a lower-level Receding Horizon game in each region $j$. Namely, we can assume company $i$'s expected profit from serving the demand in region $j$ at time interval $k$ is given by~\eqref{eq:16}, where $W_k^{j,\text{lo}}\in\R_+$ represents the time-varying total potential profit in that region. The cost of charging vehicles in region $j$ at time interval $k$ is given by $\mathbf{c}_{i,k}^j(x_{i,k}^{j,\text{lo}}, \mathbf{x}_{-i,k}^{j,\text{lo}})=(\mathbf{u}_{i,k}^{j})^T\pi_k^j$, where the marginal price vector $\pi_k^j\in\R^d$ in future liberalized markets is determined by consumer demand and is given by $\pi_k^j\defineas\alpha_k\sum_{i\in\mc I} \mathbf{u}_{i,k}^j+r_k$ for some price-scaling parameters $\alpha_k$ and $r_k$. Under this framework, following a similar reasoning as in Proposition~\ref{prop:2}, we can establish the existence and uniqueness of the Nash equilibrium over the horizon $T=9$. 

\subsection{Numerical results}\label{subsec:6b}

As discussed in Section~\ref{subsec:5b}, it is easy to verify that the upper-level Blotto game admits a unique Nash equilibrium. At the same time, finding the system-optimal strategy $\overline{\mathbf{x}}^{\text{SO}}\in\mc X$ simplifies to solving a concave optimization problem given by~\eqref{eq:xs}. If we now define the welfare of the ride-hailing market as $\text{Welf}(\mathbf{x})\defineas\sum_{i\in\mc I} \mathbf{u}_i(\mathbf{x}_i,\mathbf{x}_{-i})$, we can then explore how the Price of Anarchy (PoA) behaves numerically. The PoA measures how much efficiency is lost when decision-making is decentralized, and it is defined as the ratio between the welfare achieved under the optimal centralized solution and that at the Nash equilibrium:
\begin{equation}
    \textbf{PoA}\defineas\frac{\text{Welf}(\overline{\mathbf{x}}^{\text{SO}})}{\text{Welf}(\overline{\mathbf{x}}^{\text{NE}})}.
\end{equation}

In our setup, we consider fleet sizes of $\mathbf{R}=[200,500,1000]$ and define the region-specific average profit from serving demand as the vector $W^{\text{up}}=[220,100,50,35]\cdot 10^3$. We then examine the impact of varying the average expected cost of charging a vehicle, modeled as $\beta^{\text{up}}=[12,9,6,3]\cdot\theta$, where $\theta\in[0.1,12]$.
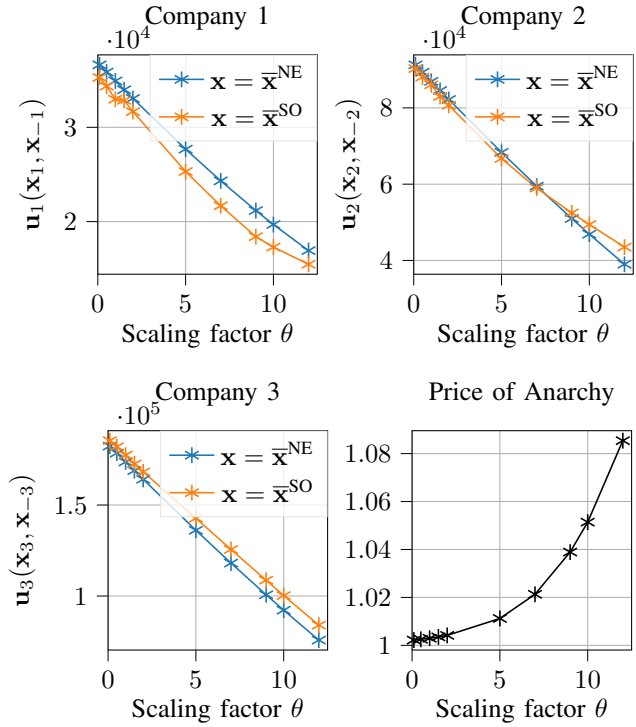
\begin {figure}
\centering
\begin{tikzpicture}[scale=1.0]

    \node[scale=1] (pic1) at (-3, 0.0) {
\begin{tikzpicture}

\definecolor{color0}{rgb}{0.12156862745098,0.466666666666667,0.705882352941177}
\definecolor{color1}{rgb}{1,0.498039215686275,0.0549019607843137}

\begin{axis}[
legend cell align={right},
legend style={fill opacity=0.8, draw opacity=1, text opacity=1, draw=white!90!black,at={(1,1)}},
title = {Company 1},
tick align=outside,
tick pos=left,
x grid style={white!69.0196078431373!black},
xlabel={Scaling factor $\theta$},
xmajorgrids,
xmin=0, xmax=12.5,
xtick style={color=black},
y grid style={white!69.0196078431373!black},
ylabel={$\mathbf{u}_1(\mathbf{x}_1,\mathbf{x}_{-1})$},
ymajorgrids,
ymin=14416.3936627212, ymax=37682.9777307375,
ytick style={color=black},
width=4.5cm,
height=4.5cm,
ylabel near ticks
]
\addplot [semithick, color0, mark=asterisk, mark size=3, mark options={solid}]
table {%
0.1 36625.4057276459
0.5 35864.5103309589
1 34920.2151631688
1.5 33983.8182092021
2 33055.6748784195
5 27682.4282633037
7 24319.9770914997
9 21171.8555219345
10 19689.0166951223
12 16924.8829523175
};
\addlegendentry{$\mathbf{x}=\overline{\mathbf{x}}^{\text{NE}}$}
\addplot [semithick, color1, mark=asterisk, mark size=3, mark options={solid}]
table {%
0.1 35292.7305957998
0.5 34358.1007484101
1 32975.5692495349
1.5 32859.0608608616
2 31664.798100082
5 25308.5253616313
7 21674.8305230549
9 18400.4891451362
10 17292.3922247474
12 15473.9656658129
};
\addlegendentry{$\mathbf{x}=\overline{\mathbf{x}}^{\text{SO}}$}
\end{axis}

\end{tikzpicture}};

    \node[scale=1] (pic2) at (1.2, 0.0) {
\begin{tikzpicture}

\definecolor{color0}{rgb}{0.12156862745098,0.466666666666667,0.705882352941177}
\definecolor{color1}{rgb}{1,0.498039215686275,0.0549019607843137}

\begin{axis}[
legend cell align={right},
legend style={fill opacity=0.8, draw opacity=1, text opacity=1, draw=white!90!black,at={(1,1)}},
title = {Company 2},
tick align=outside,
tick pos=left,
x grid style={white!69.0196078431373!black},
xlabel={Scaling factor $\theta$},
xmajorgrids,
xmin=0, xmax=12.5,
xtick style={color=black},
y grid style={white!69.0196078431373!black},
ylabel={$\mathbf{u}_2(\mathbf{x}_2,\mathbf{x}_{-2})$},
ymajorgrids,
ymin=36386.1094834924, ymax=93899.7614004233,
ytick style={color=black},
width=4.5cm,
height=4.5cm,
ylabel near ticks
]
\addplot [semithick, color0, mark=asterisk, mark size=3, mark options={solid}]
table {%
0.1 91285.5044951083
0.5 89356.5413240504
1 86958.237646569
1.5 84574.7207299112
2 82206.4948189085
5 68349.0879948816
7 59490.9752124526
9 50987.9702810711
10 46884.2036969655
12 39000.3663888074
};
\addlegendentry{$\mathbf{x}=\overline{\mathbf{x}}^{\text{NE}}$}
\addplot [semithick, color1, mark=asterisk, mark size=3, mark options={solid}]
table {%
0.1 90312.0571103662
0.5 88081.8858828154
1 86052.447466184
1.5 82980.883695676
2 80668.2569125893
5 66723.7249275096
7 58974.2960646093
9 52472.331279225
10 49424.7868756241
12 43502.7701722315
};
\addlegendentry{$\mathbf{x}=\overline{\mathbf{x}}^{\text{SO}}$}
\end{axis}

\end{tikzpicture}};

    \node[scale=1] (pic3) at (-3, -5) {
\begin{tikzpicture}

\definecolor{color0}{rgb}{0.12156862745098,0.466666666666667,0.705882352941177}
\definecolor{color1}{rgb}{1,0.498039215686275,0.0549019607843137}

\begin{axis}[
legend cell align={right},
legend style={fill opacity=0.8, draw opacity=1, text opacity=1, draw=white!90!black,at={(1,1)}},
title = {Company 3},
tick align=outside,
tick pos=left,
x grid style={white!69.0196078431373!black},
xlabel={Scaling factor $\theta$},
xmajorgrids,
xmin=0, xmax=12.5,
xtick style={color=black},
y grid style={white!69.0196078431373!black},
ylabel={$\mathbf{u}_3(\mathbf{x}_3,\mathbf{x}_{-3})$},
ymajorgrids,
ymin=70313.7096026833, ymax=190854.540103011,
ytick style={color=black},
width=4.5cm,
height=4.5cm,
ylabel near ticks
]
\addplot [semithick, color0, mark=asterisk, mark size=3, mark options={solid}]
table {%
0.1 182385.66909375
0.5 178509.926297578
1 173688.275101959
1.5 168892.891578892
2 164124.52803237
5 136126.85418262
7 118109.305374998
9 100681.494831697
10 92209.5153075427
12 75792.8382617891
};
\addlegendentry{$\mathbf{x}=\overline{\mathbf{x}}^{\text{NE}}$}
\addplot [semithick, color1, mark=asterisk, mark size=3, mark options={solid}]
table {%
0.1 185375.411443906
0.5 182054.903495996
1 177417.047413081
1.5 172621.440800629
2 168214.297450228
5 142729.612151457
7 125572.830073955
9 108694.525060056
10 100227.304103338
12 83990.4761329423
};
\addlegendentry{$\mathbf{x}=\overline{\mathbf{x}}^{\text{SO}}$}
\end{axis}

\end{tikzpicture}};

    \node[scale=1] (pic4) at (1.25, -5) {
\begin{tikzpicture}

\definecolor{color0}{rgb}{0.12156862745098,0.466666666666667,0.705882352941177}

\begin{axis}[
legend style={fill opacity=0.8, draw opacity=1, text opacity=1, draw=white!80!black},
title={Price of Anarchy},
tick align=outside,
tick pos=left,
x grid style={white!69.0196078431373!black},
xlabel={Scaling factor $\theta$},
xmajorgrids,
xmin=0, xmax=12.5,
xtick style={color=black},
y grid style={white!69.0196078431373!black},
ymajorgrids,
ymin=0.998043121686559, ymax=1.08956303025478,
ytick style={color=black},
width=4.5cm,
height=4.5cm,
ylabel near ticks
]
\addplot [semithick, black, mark=asterisk, mark size=3, mark options={solid}, forget plot]
table {%
0.1 1.00220311753057
0.5 1.00251509470579
1 1.00297170193448
1.5 1.003513479955
2 1.00415429489891
5 1.01121429304853
7 1.02130394954977
9 1.03891444953681
10 1.05140198314456
12 1.08540303441077
};
\end{axis}

\end{tikzpicture}};
    



    
\end{tikzpicture}
    \caption{\unboldmath The evolution of companies' total profits and the price of anarchy for different values of the scaling parameter $\theta\in[0.1, 12]$.} 
\label{fig:imgPoA}
\end{figure}
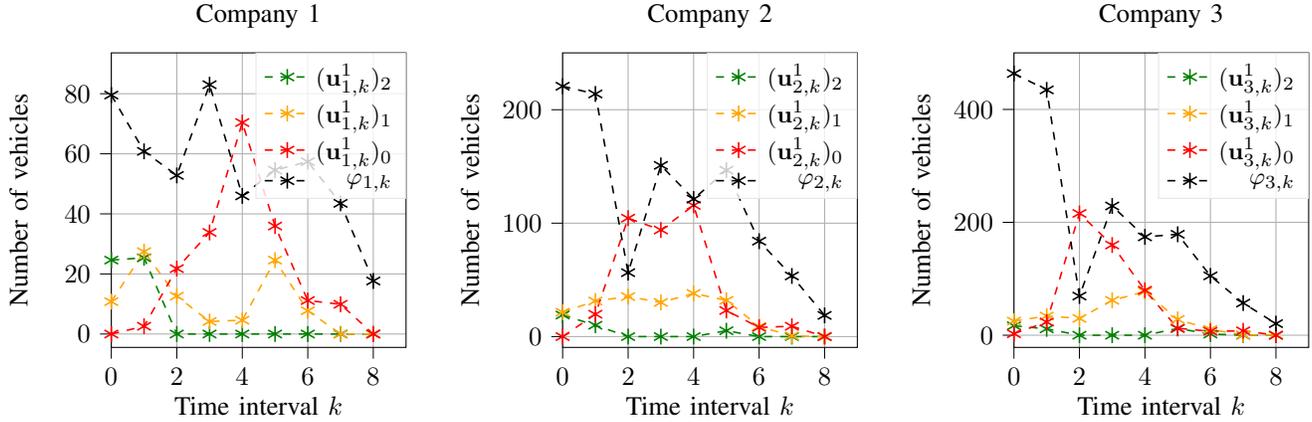
\begin {figure*}
\centering

\begin{tikzpicture}[scale=1.0]

    \node[scale=1] (pic1) at (0.0, 0.0) {
\begin{tikzpicture}

\definecolor{color0}{rgb}{1,0.647058823529412,0}

\begin{axis}[
legend cell align={right},
legend style={fill opacity=0.8, draw opacity=1, text opacity=1, draw=white!90!black,at={(1,1)}, font=\small},
title = {Company 2},
tick align=outside,
tick pos=left,
x grid style={white!69.0196078431373!black},
xlabel={Time interval $k$},
xmajorgrids,
xmin=0, xmax=9,
xtick style={color=black},
y grid style={white!69.0196078431373!black},
ylabel={Number of vehicles},
ymajorgrids,
ymin=-9.52908110062297, ymax=250.110703113082,
ytick style={color=black},
width=5.5cm,
height=5.5cm
]
\addplot [semithick, green!50.1960784313725!black, dashed, mark=asterisk, mark size=3, mark options={solid}]
table {%
0 19.0169360071708
1 10.0197459946374
2 6.77156004880419e-05
3 -4.79557787143899e-22
4 3.06177238841084e-22
5 5.17742357329761
6 1.17807283083537e-22
7 5.42008072290985e-23
8 3.98202740129248e-23
};
\addlegendentry{$(\mathbf{u}_{2,k}^1)_2$}
\addplot [semithick, color0, dashed, mark=asterisk, mark size=3, mark options={solid}]
table {%
0 21.5197931017729
1 31.353549869514
2 35.2628185479431
3 30.1190617618361
4 38.0696692863169
5 31.9586019466574
6 8.90747081482607
7 0.650877504742634
8 3.98202740129248e-23
};
\addlegendentry{$(\mathbf{u}_{2,k}^1)_1$}
\addplot [semithick, red, dashed, mark=asterisk, mark size=3, mark options={solid}]
table {%
0 1.1357962200688e-22
1 19.7302068982271
2 104.411929219239
3 93.9519563686666
4 115.666095605954
5 23.0019221793229
6 8.06479441725923
7 9.13154880454992
8 0
};
\addlegendentry{$(\mathbf{u}_{2,k}^1)_0$}
\addplot [semithick, black, dashed, mark=asterisk, mark size=3, mark options={solid}]
table {%
0 220.713270891056
1 213.896497237622
2 56.3575996131412
3 150.928981869497
4 121.264235107729
5 146.718801134152
6 84.1227225974711
7 53.4574132472976
8 18.6898971241186
};
\addlegendentry{$\varphi_{2,k}$}
\end{axis}

\end{tikzpicture}};
    
    \node[scale=1] (pic2) at (-6, 0.0) {
\begin{tikzpicture}

\definecolor{color0}{rgb}{1,0.647058823529412,0}

\begin{axis}[
legend cell align={right},
legend style={fill opacity=0.8, draw opacity=1, text opacity=1, draw=white!90!black,at={(1,1)}, font=\small},
title = {Company 1},
tick align=outside,
tick pos=left,
x grid style={white!69.0196078431373!black},
xlabel={Time interval $k$},
xmajorgrids,
xmin=0, xmax=9,
xtick style={color=black},
y grid style={white!69.0196078431373!black},
ylabel={Number of vehicles},
ymajorgrids,
ymin=-4.45744084811315, ymax=93.6062578103761,
ytick style={color=black},
width=5.5cm,
height=5.5cm
]
\addplot [semithick, green!50.1960784313725!black, dashed, mark=asterisk, mark size=3, mark options={solid}]
table {%
0 24.6106479579086
1 25.379238009602
2 -1.41812486599952e-22
3 -3.23460285622932e-22
4 -1.18032864760026e-22
5 1.02297669094142e-22
6 1.00457414796627e-22
7 2.9281213175762e-23
8 2.93702709369258e-23
};
\addlegendentry{$(\mathbf{u}_{1,k}^1)_2$}
\addplot [semithick, color0, dashed, mark=asterisk, mark size=3, mark options={solid}]
table {%
0 10.7766948049698
1 27.4281402217881
2 12.5915358264653
3 4.17786778147944
4 4.59931287765939
5 24.6059803298731
6 7.75202084507076
7 2.40983348719362e-23
8 -4.69657711700276e-25
};
\addlegendentry{$(\mathbf{u}_{1,k}^1)_1$}
\addplot [semithick, red, dashed, mark=asterisk, mark size=3, mark options={solid}]
table {%
0 3.37070121094405e-23
1 2.58343107318884
2 21.7792565543307
3 33.8218293878139
4 70.4087670042414
5 35.9437489350756
6 11.1501636259975
7 9.90253297835334
8 0
};
\addlegendentry{$(\mathbf{u}_{1,k}^1)_0$}
\addplot [semithick, black, dashed, mark=asterisk, mark size=3, mark options={solid}]
table {%
0 79.5626572371215
1 60.8193165735797
2 52.8073782313901
3 83.0003028307066
4 45.9919201180992
5 54.5799673335071
6 57.3970212975373
7 43.5081648009413
8 17.6545538234241
};
\addlegendentry{$\varphi_{1,k}$}
\end{axis}

\end{tikzpicture}};

    \node[scale=1] (pic3) at (6, 0.0) {
\begin{tikzpicture}

\definecolor{color0}{rgb}{1,0.647058823529412,0}

\begin{axis}[
legend cell align={right},
legend style={fill opacity=0.8, draw opacity=1, text opacity=1, draw=white!90!black,at={(1,1)}, font=\small},
title = {Company 3},
tick align=outside,
tick pos=left,
x grid style={white!69.0196078431373!black},
xlabel={Time interval $k$},
xmajorgrids,
xmin=0, xmax=9,
xtick style={color=black},
y grid style={white!69.0196078431373!black},
ylabel={Number of vehicles},
ymajorgrids,
ymin=-21.5799940417725, ymax=500.179874877223,
ytick style={color=black},
width=5.5cm,
height=5.5cm
]
\addplot [semithick, green!50.1960784313725!black, dashed, mark=asterisk, mark size=3, mark options={solid}]
table {%
0 16.7262024257391
1 11.2270173878109
2 1.57883361087208e-22
3 4.07705209689957e-22
4 2.9408709029071e-22
5 11.9144728712507
6 2.11085590606011
7 5.45370694785631e-23
8 6.82813798501125e-23
};
\addlegendentry{$(\mathbf{u}_{3,k}^1)_2$}
\addplot [semithick, color0, dashed, mark=asterisk, mark size=3, mark options={solid}]
table {%
0 25.0426364371077
1 34.0509183591431
2 29.7885741722
3 61.8421401145693
4 77.2050039449355
5 28.0468329354029
6 10.3666995648234
7 0.344522743889155
8 6.82813798501125e-23
};
\addlegendentry{$(\mathbf{u}_{3,k}^1)_1$}
\addplot [semithick, red, dashed, mark=asterisk, mark size=3, mark options={solid}]
table {%
0 2.65088599133669
1 23.3584298323342
2 215.731805250316
3 159.780345539672
4 79.7283904728746
5 12.3531038714567
6 6.97811046708333
7 7.73851009352054
8 0
};
\addlegendentry{$(\mathbf{u}_{3,k}^1)_0$}
\addplot [semithick, black, dashed, mark=asterisk, mark size=3, mark options={solid}]
table {%
0 463.631161137153
1 434.61558668149
2 69.3896128821241
3 228.956175054901
4 174.206055881506
5 178.814228725726
6 105.127385280912
7 56.9615930946711
8 20.5605883082932
};
\addlegendentry{$\varphi_{3,k}$}
\end{axis}

\end{tikzpicture}};
        
\end{tikzpicture}
\caption{\unboldmath Open loop control, i.e., $T=9$: The figure illustrates the optimal control input over the planning horizon, along with the corresponding auxiliary market participation for each ride-hailing company. With a slight abuse of notation, $\varphi_{i,k}=\varphi_{i,k}(\mathbf{y}_{i,k}^1,\mathbf{u}_{i,k}^1)$ denotes company $i$'s market participation at time interval $k$. Meanwhile, $(\mathbf{u}_{i,k}^1)_0\in\R_+$, $(\mathbf{u}_{i,k}^1)_1\in\R_+$, and $(\mathbf{u}_{i,k}^1)_2\in\R_+$ represent the number of vehicles with 'red,' 'yellow,' and 'green' battery levels, respectively, that are sent to recharge at time interval $k$.}

\label{fig:lower_level}
\end{figure*}
Figure~\ref{fig:imgPoA} illustrates how scaling the charging cost through different values of $\theta$ affects both the companies' profits and the resulting Price of Anarchy. As expected from~\eqref{eq:xs}, higher values of $\theta$ make the advantages of centrally optimizing the market's welfare more apparent. Additionally, the plots indicate that the system-optimal strategy tends to benefit the third company the most, while the first and second companies often achieve higher profits under the Nash Equilibrium strategy. This suggests that the size of the operating fleet could also play a role in the behavior of PoA, as the charging costs become more prominent for larger fleets. Regarding execution times, we report that Algorithm~\ref{al:1} requires approximately $\mathbf{T}_{\text{alg},1}\approx 5.4 \text{ sec}$ to compute the NE of the Blotto game for a tolerance of $\text{tol}=10^{-5}$. In contrast, the semi-analytical approach, combined with off-the-shelf root-finding procedures from the 'scipy.optimize' library achieves an execution time of just $0.3\cdot 10^{-3} \text{ sec}$ for finding the interior-point NE. When accounting for the worst-case multiplicative factor for $K=4$ and $N=3$, this results in a conservative estimate of $\mathbf{T}_{\text{alg},2}\approx 1.02 \text{ sec}$, highlighting a substantial reduction in execution time in practice. 

To illustrate lower-level management, we examine receding horizon charging scheduling problem in region $J_1$ in a simulation that spans $T_{\text{tot}}=9$ time steps. The solution to the upper-level management problem for $\theta=1$ determines the region-specific fleet sizes for each company: $x_{1,1}^{\text{up}}=121$, $x_{2,1}^{\text{up}}=275$, and $x_{3,1}^{\text{up}}=532$. The region-specific parameters $W_k^{1,\text{lo}}$, $\alpha_k$ and $\varepsilon_k$ follow the temporal profiles shown in Figure~\ref{fig:img_taccasestudy1} and each vehicle is classified into one of three battery categories: red, yellow, or green. We adopt that, at the beginning of the simulation, $85\%$ of the vehicles are green, $10\%$ are yellow, $5\%$ are red, and that the charging cost has the corresponding parameter $r_k=0$ for every $k\in\Z_{T_{\text{tot}}}$. We evaluate the performance of the receding-horizon planner under two scenarios: first, in an open-loop setting where the planning horizon is equal to the total duration, $T=T_{\text{tot}}$, and second, in a closed-loop fashion with shorter horizons of $T=6$ and $T=3$. Figure~\ref{fig:lower_level} shows market participation and the optimal number of vehicles from each battery category sent for recharging when the planner has full knowledge of the temporal evolution of the parameters. As expected, during peak hours, companies increase their market participation to maximize profits. To achieve this, they strategically recharge their vehicles during periods of lower electricity prices, balancing the need to maintain a high number of operational vehicles during peak demand with ensuring sufficient battery levels to avoid forced parking. Conversely, the end of the simulation, rising charging costs and diminishing potential profits result in fewer vehicles being sent for recharging, leading to lower market participation due to reduced vehicle availability. This outcome is expected, as there are no parking costs, and with the time horizon nearing completion, the incentive to proactively charge vehicles diminishes. Similar trends, often referred to as 'end-of-day' effects, have been documented in the literature~\cite{6994293, 9354436, ESTRELLA2019126}.  

\begin{figure}[!t]
    \centering
\begin{tikzpicture}

\definecolor{color0}{rgb}{0.12156862745098,0.466666666666667,0.705882352941177}
\definecolor{color1}{rgb}{1,0.498039215686275,0.0549019607843137}
\definecolor{color2}{rgb}{0.172549019607843,0.627450980392157,0.172549019607843}

\begin{axis}[
legend cell align={right},
legend style={fill opacity=0.8, draw opacity=1, text opacity=1, draw=white!90!black,at={(1,1)}, font=\small},
tick align=outside,
tick pos=left,
x grid style={white!69.0196078431373!black},
xlabel={Time interval $k$},
xmajorgrids,
xmin=0, xmax=9,
xtick style={color=black},
y grid style={white!69.0196078431373!black},
ylabel={$\Psi_k(\Phi_{k,\mathbf{x}})$},
ymajorgrids,
ymin=-1462.16771685889, ymax=31466.4124070771,
ytick style={color=black},
height = 6cm
]
\addplot [semithick, color0, const plot mark left, dashed, mark=asterisk, mark size=3, mark options={solid}]
table {%
0 33.1277537977635
1 272.62627938088
2 4282.66882284131
3 5903.07780964565
4 4901.46608383903
5 2448.69595686893
6 1396.93990906996
7 751.145566394276
8 733.729100071983
9 733.729100071983
};
\addlegendentry{$T=9$}
\addplot [semithick, color1, const plot mark left, dashed, mark=asterisk, mark size=3, mark options={solid}]
table {%
0 32.8657728813616
1 273.491235259308
2 4125.71059179307
3 5766.48539036495
4 4230.80565612
5 2551.5420199289
6 4540.20425874033
7 1716.36189265562
8 3892.60222999494
9 3892.60222999494
};
\addlegendentry{$T=6$}
\addplot [semithick, color2, const plot mark left, dashed, mark=asterisk, mark size=3, mark options={solid}]
table {%
0 32.3036193191259
1 245.99155643871
2 3513.06983396404
3 29450.7339194516
4 20961.6019671666
5 28135.4061494466
6 20000
7 1901.02578217268
8 796.948490228198
9 796.948490228198
};
\addlegendentry{$T=3$}
\end{axis}

\end{tikzpicture}
    \caption{\unboldmath Evolution of the stage profit loss $\Psi_k(\Phi_{k,\mathbf{x}})$ for different planning horizons $T\in\{3,6,9\}$.}
    \label{fig:horizon}
\end{figure}
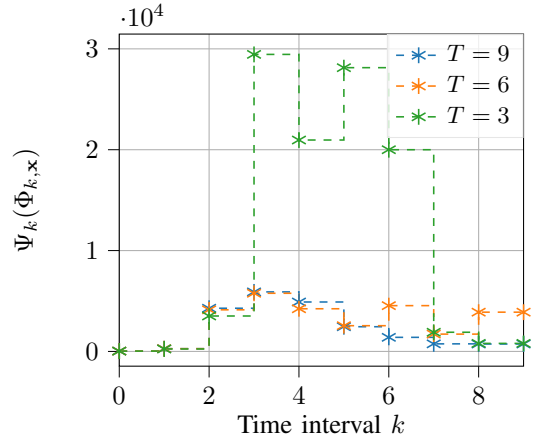

The impact of different planning horizon lengths is illustrated in Figure~\ref{fig:horizon} and Table~\ref{tab:horizon}. As expected, a longer planning horizon enables companies to better anticipate fluctuations in demand and charging costs, leading to more effective charging strategies and, ultimately, higher overall profits.

Notably, when $T=3$, companies are unprepared for peak demand periods, resulting in significant stage profit losses. In contrast, horizons of $T=6$ and $T=9$ yield substantially better performance, with the primary difference in profit loss between these cases arising toward the end of the simulation period, where $T=6$ shows slightly weaker performance. These findings are further supported by the numerical results in Table~\ref{tab:horizon}, which show that all three companies experience an increase in total profit as the planning horizon extends, thereby reducing profit losses from insufficient market participation. Interestingly, Companies 2 and 3, which operate larger fleets, see a more significant rise in total profit for $T\in\{3,6\}$ compared to Company 1. However, for $T=3$, Company 1, despite having the smallest fleet, attains the highest total profit. 
\begin{table}[!t]
\begin{center}
 \renewcommand{\arraystretch}{1.3}
 \caption{Attained profits for different planning horizons}\vspace{1ex}
 \label{tab:horizon}
 \begin{tabular}{c|c|c|c|c}
 Horizon  & Company 1 & Company 2 & Company 3 & Lost profit \\ 
 \hline\hline
 $T=3$ &  19312 & 17497 & 18872 & 105037    \\
 $T=6$ &  21507 & 41007  & 60476 & 27130   \\
 $T=9$ &  22023 & 41429  & 65532 & 20723   \\
\hline 
\end{tabular}
\end{center}
\vspace{-2em}
\end{table}

\section{Conclusion}\label{sec:7}

This paper introduced a new class of resource-splitting games with lossy stage profits, providing a framework that better reflects real-world market dynamics, where profitability depends not only on resource investment but also on sufficient player participation. By incorporating a structured model of profit loss and demonstrating how a Tullock-like payoff structure naturally emerges under weighted proportional fair resource allocation, we established a theoretical foundation for analyzing both centralized and self-interested strategies. Our analysis established sufficient conditions for the existence and uniqueness of Nash equilibria and introduced an iterative, semi-decentralized method for computing them. Furthermore, we showed that our framework generalizes well-known game-theoretic models, such as Receding Horizon and Blotto games, and developed a semi-analytical approach for computing equilibrium strategies in the latter. Through a case study in smart mobility, we validated the practical applicability of our model and its potential to inform decision-making in competitive multi-agent markets. We illustrated how the Price of Anarchy in the Blotto setup with the standard notion of system welfare behaves numerically and provided insights on the effects of varying the length of the planning horizon in the Receding horizon setup. Future work will explore extensions to dynamic and learning-based settings, where agents adapt strategies over time based on observed market responses, further bridging the gap between theoretical game models and real-world applications.

\section*{References}

\bibliographystyle{IEEEtran}
\bibliography{references.bib}

\begin{IEEEbiography}[{\includegraphics[width=1in,height=1.25in,clip,keepaspectratio]{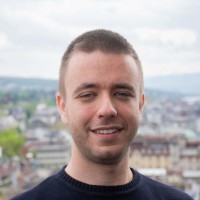}}]{Marko Maljkovic} received his B.Sc. in Electrical and Computer Engineering from the School of Electrical Engineering, University of Belgrade, and his M.Sc. in Robotics, Systems, and Control from ETH Zurich in 2018 and 2021, respectively. He is currently a doctoral student at the Urban Transport Systems Laboratory (LUTS) at EPFL. His research interests lie in game-theoretic and learning-based mechanism design and control for multi-agent systems, with applications in future mobility systems.
\end{IEEEbiography}

\begin{IEEEbiography}[{\includegraphics[width=1in,height=1.25in,clip,keepaspectratio]{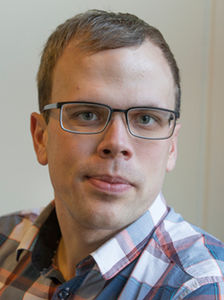}}]{Gustav Nilsson} received his M.Sc.~in Engineering Physics and Ph.D.~in Automatic Control from Lund University in 2013 and 2019, respectively. He is currently a Postdoctoral researcher att University of Trento. Prior to his current position, he has been a Postdoctoral Researcher at EPFL and a Postdoctoral Associate at GeorgiaTech, GA, USA. His primary research interest lies in modeling and control of dynamical flow networks with applications in transportation networks.
\end{IEEEbiography}

\begin{IEEEbiography}[{\includegraphics[width=1in,height=1.25in,clip,keepaspectratio]{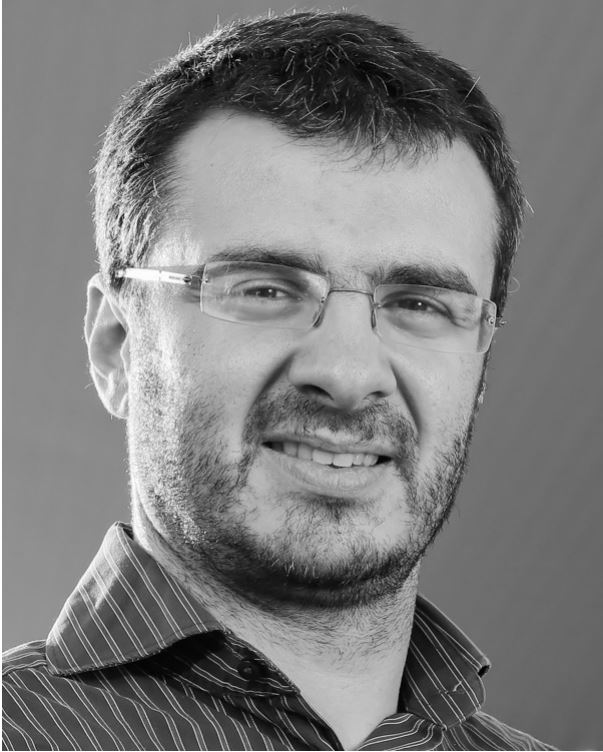}}]{Nikolas Geroliminis} is a Full Professor at EPFL and the head of the Urban Transport Systems Laboratory (LUTS). Before joining EPFL he was an Assistant Professor on the faculty of the Department of Civil Engineering at the University of Minnesota. He has a diploma in Civil Engineering from the National Technical University of Athens (NTUA) and a M.Sc. and Ph.D. in civil engineering from University of California, Berkeley. His research interests focus primarily on urban transportation systems, traffic flow theory and control, public transportation and on-demand transport, car sharing, Optimization and Large Scale Networks. Among his recent initiatives is the creation of an open-science large-scale dataset of naturalistic urban trajectories of half a million vehicles that have been collected by one-of-a-kind experiment by a swarm of drones (https://open-traffic.epfl.ch). Among other editorial responsibilities, he is currently the Editor-In-Chief of Transportation Research part C: Emerging Technologies.
\end{IEEEbiography}

\end{document}